\documentclass[structabstract]{aa}  
\usepackage{graphicx}
\usepackage{natbib}
\usepackage{txfonts}
\newcommand{\kmps}{\ensuremath{\mathrm{km\,s^{-1}}}}
\newcommand{\mps}{\ensuremath{\mathrm{m\,s^{-1}}}}

\newcommand{\hour}{\ensuremath{^\mathrm{h}}}
\newcommand{\minute}{\ensuremath{^\mathrm{m}}}

\begin{document}
   \title{High-resolution ammonia mapping of the very young protostellar 
          core Chamaeleon-MMS1
             \thanks{Based on observations obtained with the 64-m Parkes
            radio telescope and the Australia Telescope Compact Array
            (ATCA). The Parkes radio telescope and ATCA are part of
            the Australia Telescope National Facility which is funded
            by the Commonwealth of Australia for operation as a
            National Facility managed by CSIRO.}}

   \author{M.S. V\"ais\"al\"a \inst{1} \and  J. Harju \inst{2,1} \and 
    M.J. Mantere \inst{3,1} \and O. Miettinen  \inst{1} \and R.S. Sault \inst{4}
          \and C.M. Walmsley \inst{5,6}
          \and J.B. Whiteoak \inst{4}
	      }  

   \institute{Department of Physics, P.O. Box 64, FI-00014 University of 
    Helsinki, Finland
   \and 
    Finnish Centre for Astronomy with ESO (FINCA), University of Turku,
    V\"{a}is\"{a}l\"{a}ntie 20, FI-21500 Piikki\"{o}, Finland
     \and
      Aalto University, Department of Information and Computer Science,
      PO Box 15400, FI-00076 Aalto, Finland  
    \and  Australia Telescope National Facility, CSIRO, PO Box 76, Epping, 
            NSW 1710, Australia  
     \and
    Osservatorio Astrofisico di Arcetri, Largo E. Fermi 5, I-50125 Firenze, 
    Italy
    \and
    Dublin Institute for Advanced Studies (DIAS), 31 Fitzwilliam Place, 
    Dublin 2, Ireland 
     }

   \date{Received XX XX, 2013; accepted XX XX, 2013}

 
  \abstract
   {}
   {The aim of this study is to investigate the structure and 
    kinematics of the nearby candidate first hydrostatic core Cha-MMS1.}
  {Cha-MMS1 was mapped in the NH$_3(1,1)$ line and the 1.2 cm
    continuum using the Australia Telescope Compact Array, ATCA.  The
    angular resolution of the ATCA observations is $7\arcsec$ ($\sim 1000$ AU),
    and the velocity resolution is 50 m~s$^{-1}$. The core was also 
    mapped with the 64-m Parkes telesope in the NH$_3(1,1)$ and
    $(2,2)$ lines. Observations from \textit{Herschel Space
      Observatory} and \textit{Spitzer Space telescope} were used to
    help interpretation. The ammonia spectra were analysed using
    Gaussian fits to the hyperfine structure. A two-layer model was
    applied in the central parts of the core where the ATCA spectra 
    show signs of self-absorption.}
  { A compact high column density core with a steep velocity
      gradient ($\sim 20$ \kmps~ pc$^{-1}$) is detected in ammonia. We
      derive a high gas density ($\sim 10^{6}\,{\rm cm}^{-3}$) in this
      region, and a fractional ammonia abundance compatible with
      determinations towards other dense cores ($\sim 10^{-8}$). This
      suggests that the age of the high density core is comparable to
      the freeze-out timescale of ammonia in these conditions, of the
      order of $10^4$ years.
    The direction of the velocity gradient agrees with previous
    single-dish observations, and the overall velocity distribution can
    be interpreted as rotation. The rotation axis goes through the
    position of a compact far-infrared source detected by
    \textit{Spitzer} and \textit{Herschel}. The specific angular
    momentum of the core, $\sim 10^{-3}$ \kmps\,pc, is typical for
    protostellar envelopes.  A string of 1.2 cm continuum sources is
    tentatively detected near the rotation axis. The ammonia
      spectra suggest the presence of warm embedded gas in its
      vicinity. An hourglass-shaped structure is seen in ammonia at
    the cloud's average LSR velocity, also aligned with the rotation
    axis. Although this structure resembles a pair of outflow lobes
    the ammonia spectra show no indications of shocked gas. }
 {
   The observed ammonia structure mainly delineates the inner envelope
   around the central source. The velocity gradient is likely to
   originate in the angular momentum of the contracting core, although
   influence of the outflow from the neighbouring young star IRS4 is
   possibly visible on one side of the core. The tentative
   continuum detection and the indications of a warm background component near 
   the rotation axis suggest that the core contains a deeply embedded outflow 
   which may have been missed in previous single-dish CO surveys owing to beam 
   dilution.
   }


   \maketitle
%

\section{Introduction}

\label{Introduction}

The characteristics of a stellar system in formation are determined by
the mass and angular momentum distributions of the place of its
origin, a collapsing dense molecular cloud core. According to the
current view the early phases of the low-mass protostellar collapse can be
roughly divided into two stages.  The collapse leads initially to the
formation of the so-called first hydrostatic core (FHSC,
\citealt{1969MNRAS.145..271L}) that, depending on the core rotation,
can resemble a disk-like structure. Soon thereafter (within a few
thousand years) the dissociation of molecular hydrogen allows the core
material to collapse further into the system equator forming a
protostar and an accretion disk around it
(e.g., \citealt{2008ApJ...676.1088M}).  The system looses mass and
angular momentum through outflows, which can be driven by collimated
jets or magnetocentrifugally accelerated disk winds (e.g.,
\citealt{2007prpl.conf..277P}).

The situation can be complicated if the magnetic field and the
rotation axis are misaligned, or if the core collapses under an
external influence.  In view of the fact that most stars form in
clusters with various disturbing elements, it is evident that real
situations are not simple, and both theoretical modelling and
high-resolution observational studies in different environments are
needed.

In this paper, we concentrate on gas kinematics in the nearby dense
core Cha-MMS1, as probed by ammonia observations with the Australia
Telescope Compact Array (ATCA). The core was discovered in the
  $\lambda=1.3$ mm survey by \cite{1996A&A...314..258R}. It is
associated with the reflection nebula Cederblad 110 and a small
cluster of young stellar objects (YSOs) located in the centre of the
Chamaeleon I (Cha I) cloud complex (e.g.,
\citealt{1991MNRAS.251..303P}; \citealt{2001A&A...376..907P}). The
distance to Cha I is $\sim 160$ pc \citep{1997A&A...327.1194W}.
Amongst star-forming cores Cha-MMS1 is an interesting specimen in two
respects.  Firstly, owing to the \textit{Spitzer} detection at 24 and
70 $\mu$m and the lack of high-velocity line emission, the source has
been suggested to represent a FHSC (\citealt{2006A&A...454L..51B};
\citealt{2011A&A...527A.145B}; \citealt{2012ApJ...744..131C};
\citealt{2013A&A...557A..98T}).  Secondly, the core is obviously hit
by molecular outflow from the neighbouring Class I protostar Ced110
IRS4 (\citealt{2007ApJ...664..964H}; \citealt{2011ApJ...743..108L}),
making it therefore a potential example of externally triggered
collapse.

The main motivation of the present study is to use the gas motions 
to study the evolutionary stage of the core, and to examine the
role of the suggested external momentum input. It turns out that the
high spectral resolution combined with a reasonably good spatial
resolution provided by ATCA reveal a peculiar rotating structure,
probably resulting from spin-up during the core contraction, but which
has a slight indication of external influence on one side of the core.
We also find indirect evidence for an embedded outflow. 

In Sect.~\ref{obsreduct} of this paper, we describe the observing methods and 
the data reduction, and in Sect.~\ref{results}, we present the results of
the observations.  In Sect.~\ref{discussion}, we discuss the core kinematics 
in the light of previous observations and the predictions 
of recent MHD models for a collapsing rotating core. Finally, in 
Sect.~\ref{conclusions}, we present our conclusions.


\section{Observations and data reduction}

\label{obsreduct}

\subsection{Parkes mapping}
\label{Parkes_mapping}

The dense clump around the millimetre source Cha-MMS1 was mapped in
the $(J,K)=(1,1)$ and $(2,2)$ inversion lines of NH$_3$ at
$\lambda=1.3$ cm using the Parkes 64-m telescope in January 1991. The
HPBW of the telescope is $80\hbox{$^{\prime\prime}$ }$ at the
wavelength used. The single channel maser receiver was connected to a
1\,024 channel digital autocorrelator which was split into two bands
512 channels each to record the $(1,\,1)$ and $(2,\,2)$ spectra
simultaneously. The width of each spectral band was 5 MHz, and
  the velocity resolution was 0.12 \kmps. The observations were done
in the position switching mode. The mapped area is about
$5\arcmin\times5\arcmin$.  The (0,0) position was IRAS 11054-7706
  C \citep{1987ApJ...316..311C} (R.A. $11^{\rm h}06^{\rm m}53\fs0$,
  Dec. $-77\degr22\arcmin47\arcsec$ (J2000)), later identified as Ced
  110 IRS4 \citep{1991MNRAS.251..303P}, and the map was extended
  towards southwest where strong ammonia emission was found. The
  spacing between map positions was generally $1\arcmin$, but a spacing of
  $30\arcsec$ was used in the vicinity of the ammonia maximum. At the
  time of the observations the millimetre source was not yet
  discovered. The calibration was checked by observing standard
extragalactic sources at different elevations, and by comparing the
ammonia line intensities towards some strong galactic sources also
visible from the Effelsberg 100-m telescope in Germany, taking the
different beam sizes into account.

\begin{figure}[ht]
   \centering
   \includegraphics[width=9.0cm,angle=0]{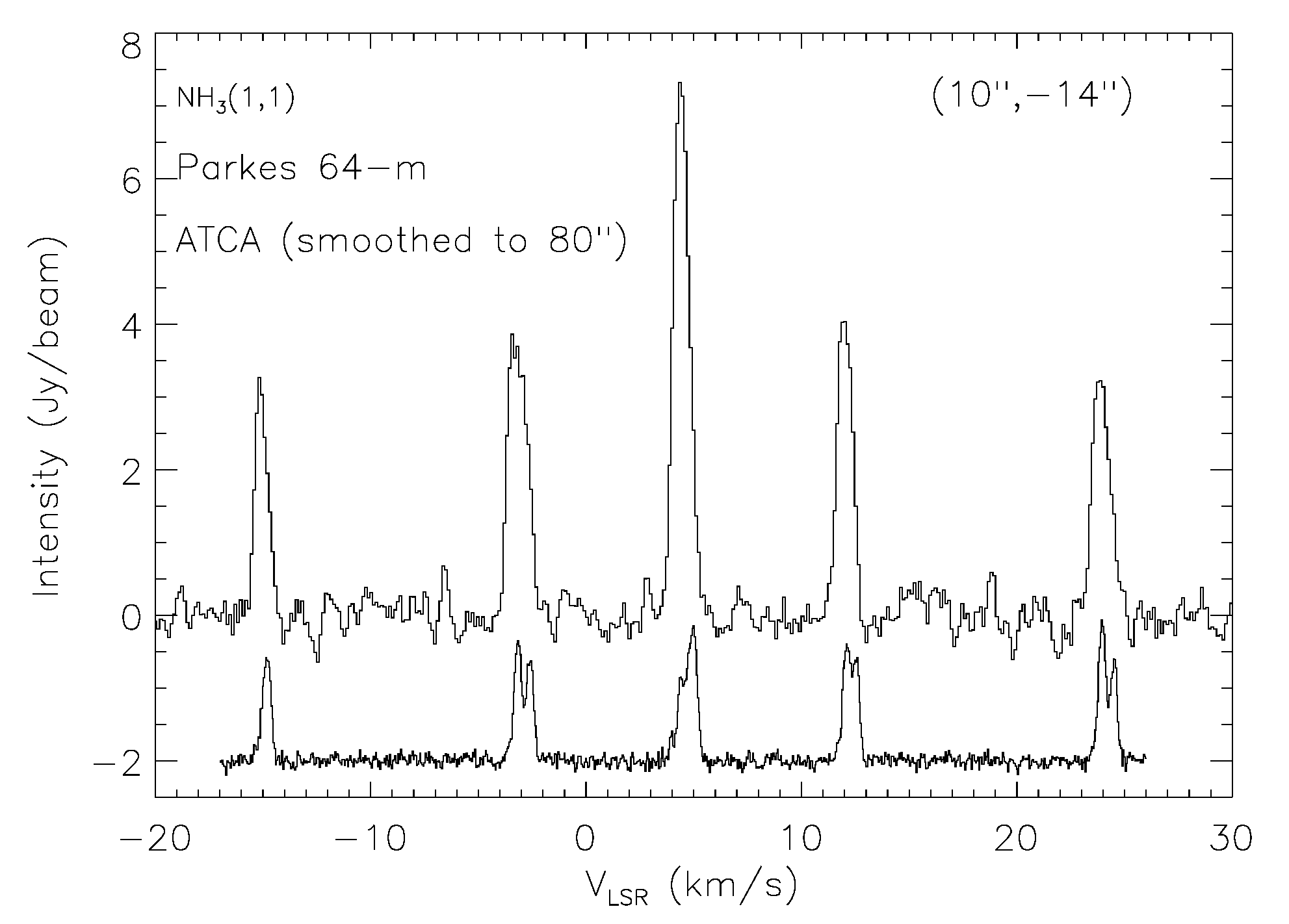}

   \caption{The NH$_3(1,1)$ inversion line spectrum towards the ammonia
     maximum as observed with the Parkes 64-m telescope (upper
     spectrum), at the offset $(+10\arcsec,-14\arcsec)$ from Cha-MMS1a
     \citep{1996A&A...314..258R}. The corresponding ATCA spectrum
     ``smoothed'' to an $80\arcsec$ resolution is shown in the
     bottom. The latter is produced by applying a Gaussian taper to
     the visibility data.}
\label{figure:Parkes_spectrum}
\end{figure}

The two brightest positions in both $(1,1)$ and $(2,2)$ lie at the
offsets $(+10\arcsec,-14\arcsec)$ and $(+10\arcsec,+16\arcsec)$ from
Cha-MMS1a \citep{1996A&A...314..258R} the coordinates of which are
R.A. $11^{\rm h}06^{\rm m}31\fs7$, Dec. $-77\degr23\arcmin32\arcsec$
(J2000).  The $(1,1)$ spectrum towards $(+10\arcsec,-14\arcsec)$ is shown in
Fig.~\ref{figure:Parkes_spectrum}, together with the ATCA spectrum
convolved to the same angular resolution. These spectra are presented
in the intensity unit Jy~beam$^{-1}$.  The position $(+10\arcsec,-14\arcsec)$
is in fact the same as shown in Fig.~6 of \citet{2006A&A...456.1037T}
from a different observing run.\footnote{We note that the NH$_3(1,1)$
  and $(2,2)$ spectra of Tennekes et al. are on the $T_{\rm MB}$ scale
  although the $y$-axis label of their Fig.~6 tells otherwise.} The
integrated NH$_3(1,1)$ intensity map is shown in
Fig.~\ref{figure:Herschel250}, superposed on the total H$_2$ column
density map as derived from thermal dust emission
(Sect.~\ref{Herschel_data}).  Analysis of the Parkes data is presented in 
Sect.~\ref{Parkes_results}.

\begin{figure}[ht]
   \centering
   \includegraphics[width=9.0cm,angle=0]{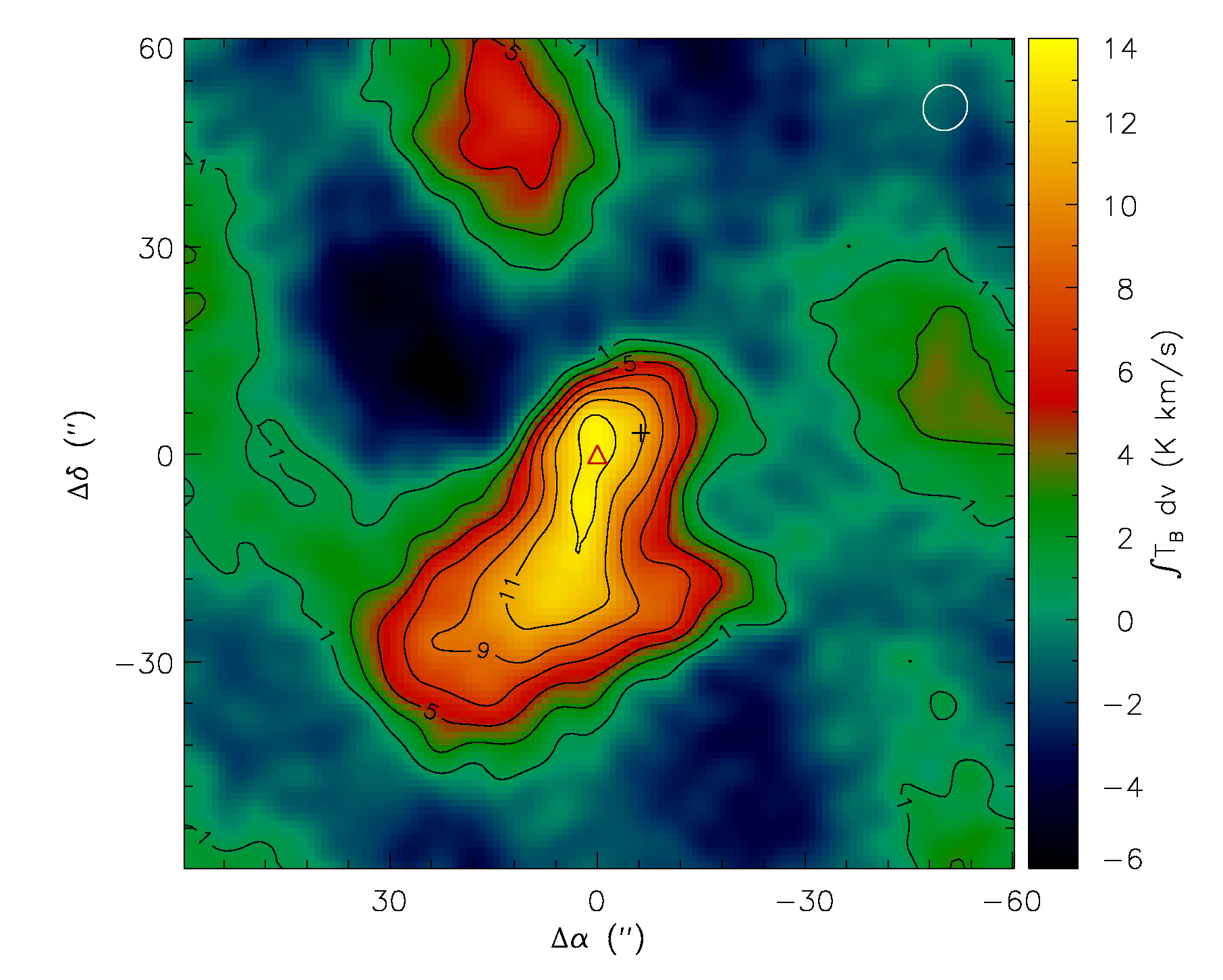}
\caption{The integrated intensity map of the NH$_3(1,1)$ satellites
  towards Cha-MMS1 observed with ATCA. The positions of Cha-MMS1a and
  the \textit{Spitzer} 24-$\mu$m maximum are marked with a plus sign and a
  triangle, respectively. The synthesised beam is indicated in the top right.}
\label{figure:Satellite_area_map}
\end{figure}


\subsection{ATCA mapping}
\label{ATCA_mapping}

\begin{table}
\caption{Observational parameters of the ATCA mapping}  
\centering
\label{table:observations}
\small
\begin{tabular}{c c c}\hline\hline
Target source&  \multicolumn{2}{c}{Cha-MMS1} \\
Phase centre  & $\alpha_{2000.0}=11\hour06\minute31\fs7$ 
              & $\delta_{2000.0}=-77\degr23\arcmin42\arcsec$ \\
Centre frequency   & \multicolumn{2}{c}{23\,692.5 MHz}   \\
Correlator config. &  \multicolumn{2}{c}{FULL\_4\_1024-128} \\

Dates         & Sep 6, 2003 &   Sep 30, 2003 \\

ATCA configuration &  EW367B & EW352 \\ 

Flux calibrator    & Jupiter  & Mars \\
                     
Bandpass calibrator & 1921-293 & 1253-055 \\
                    & $S_\nu = 13.58\pm0.03$\,Jy  & 
$S_\nu = 18.01\pm0.05$\,Jy \\
 
Phase calibrator    & \multicolumn{2}{c}{1057-797} \\
                    & \multicolumn{2}{c}{$S_{23.7 \rm GHz} = 
                       1.84\pm0.05$\,Jy} \\
                    & \multicolumn{2}{c}{Linear polarisation}\\
                    & \multicolumn{2}{c}{4.7\% (P.A.\ $-63^\circ$)} \\  

Image processing    & \multicolumn{2}{c}{Naturally weighted}        \\
Synthesised beam    & \multicolumn{2}{c}{$6\farcs9\times 6\farcs4$
                      (P.A.\ $-37^\circ$)} \\
Primary beam        & \multicolumn{2}{c}{$2\farcm4$} \\
$1\sigma$ rms noise  & \multicolumn{2}{c}{9.5 mJy\,beam$^{-1}$ or 0.5 K per 3.9 kHz channel} \\
                    &   \\ 
\hline
\end{tabular}
\label{table:obs}
\end{table}

Cha-MMS1 was observed with the ATCA, located near Narrabri, New South
Wales, Australia. The observations were made using a single
  pointing. The image covers a region of about $2\arcmin$ in diameter
  around the phase centre at $\alpha_{2000.0}=11\hour06\minute31\fs7$,
  $\delta_{2000.0}=-77\degr23\arcmin42\arcsec$, which lies $10\arcsec$
  south of the 1.3-mm dust continuum maximum Cha-MMS1a
  \citep{1996A&A...314..258R}. The observations were made in two 12
hours runs on 2003 Sep 6 and 30, using the array configurations EW367B
and EW352.  The shortest baselines of these configurations are about
46 and 31 m, respectively. The longest baselines for both
configurations are about 4\,400 m.

The phase and bandpass ca\-lib\-ra\-ti\-on sources were selected from
the ATCA Calibrator Source Catalogue (Reynolds et al. 1997).  The
phase-reference calibrator was the quasar PKS~1057--797, and this
source was used also for the reference pointing measurement in the
beginning of every fourth phase calibrator - target source cycle. The
duration of the cycle was 3 + 15 minutes.  The bandpass calibrators
were PKS~1921-293 and PKS~1253-055. The planets Jupiter and Mars were
observed to provide flux standards.  The absolute calibration was done
using the predicted visibility functions provided by the Miriad
reduction package \citep{1995ASPC...77..433S}.

The correlator was set up to record the NH$_3(1,1)$ inversion line and
the 1.3 cm continuum. The NH$_3$(1,1) line has 18 hyperfine components
concentrated on 5 groups, the so-called main group and four
satellites. The rest frequency of the line centre is about 23\,694.5
MHz, and the separation between the outermost satellites is 3.3 MHz.
The correlator set-up gave 1) a 4 MHz frequency band with 1\,024
channels (for lines), and 2) a 112 MHz band divided into 14 channels
of 8 MHz (for continuum).  The spectral resolution of the narrow band
corresponds to 50 \mps.  Both bands were centred at 23\,692.5 MHz.
The Doppler correction was done off-line in the course of data
reduction. The 1.3 cm receivers have orthogonal linearly polarised
feeds. The correlator yielded two polarisation products in the narrow
band, and all four in the wide band.  The observational parameters are
summarised in Table~\ref{table:obs}.

The weather was good on Sep 6, but at best moderate during the second
session on Sep 30. The theoretical visibility noise in the
  continuum had an rms of $\sim 0.03-0.04$ Jy on Sep 6.  On Sep 30 the
  rms noise was $\sim 0.07$ Jy in the beginning but rose above 0.1 Jy
  after 6 hours of measurements because of high humidity.  Only the
  first half of the second run was used.

The data were calibrated, inverted and cleaned using Miriad. The
inversion of the wideband visibility data was performed using all the
channel information (i.e.,\ multifrequency synthesis).  The flux
densities of the detectected sources were derived from naturally
weighted images corrected for the primary beam responses.  The noise
level at the centre of the image, and the size of the synthesised beam
are given in Table~\ref{table:observations}.

The ATCA spectrum towards $(10\arcsec,-14\arcsec)$ smoothed to the
angular resolution of Parkes, $80\arcsec$, is shown in
Fig.~\ref{figure:Parkes_spectrum} (lower curve). The smoothed image
cube has been produced by weighting (``tapering'') the visibility data
by a Gaussian in the inversion. The ATCA data observations miss a
substantial part of the flux owing to the lack of baselines shorter
than 30 m.  Comparing the NH$_3(1,1)$ satellites in the Parkes and
smoothed ATCA spectra shows that the missing flux is about 50\%.

The integrated NH$_3(1,1)$ satellite intensity map of Cha-MMS1
observed with with ATCA is shown in
Fig.~\ref{figure:Satellite_area_map}. The main group has been excluded
here because of its large optical thickness. The analysis of the
ATCA wide-band images is described in Sect.~\ref{widebandimages}, and
the spectra are analysed in Sect.~\ref{ATCA_spectral_images}.

\subsection{Herschel data}
\label{Herschel_data}

\begin{figure*}[ht]
\unitlength=1mm
\begin{picture}(210,70)
\put(-5,0){\begin{picture}(0,0)
\includegraphics[width=6.7cm,angle=0]{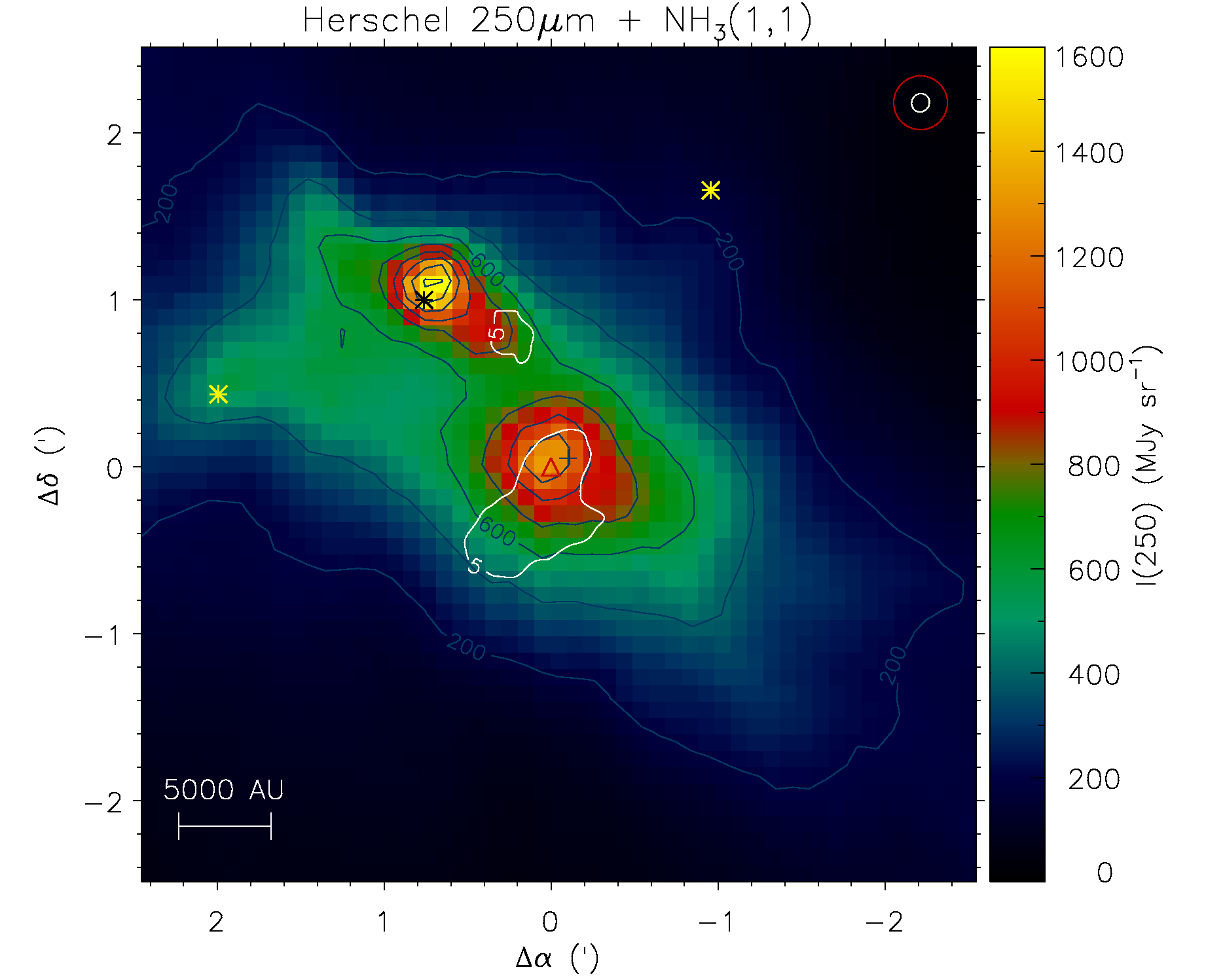} 
\end{picture}}
\put(125,0){\begin{picture}(0,0)
\includegraphics[width=6.4cm,angle=0]{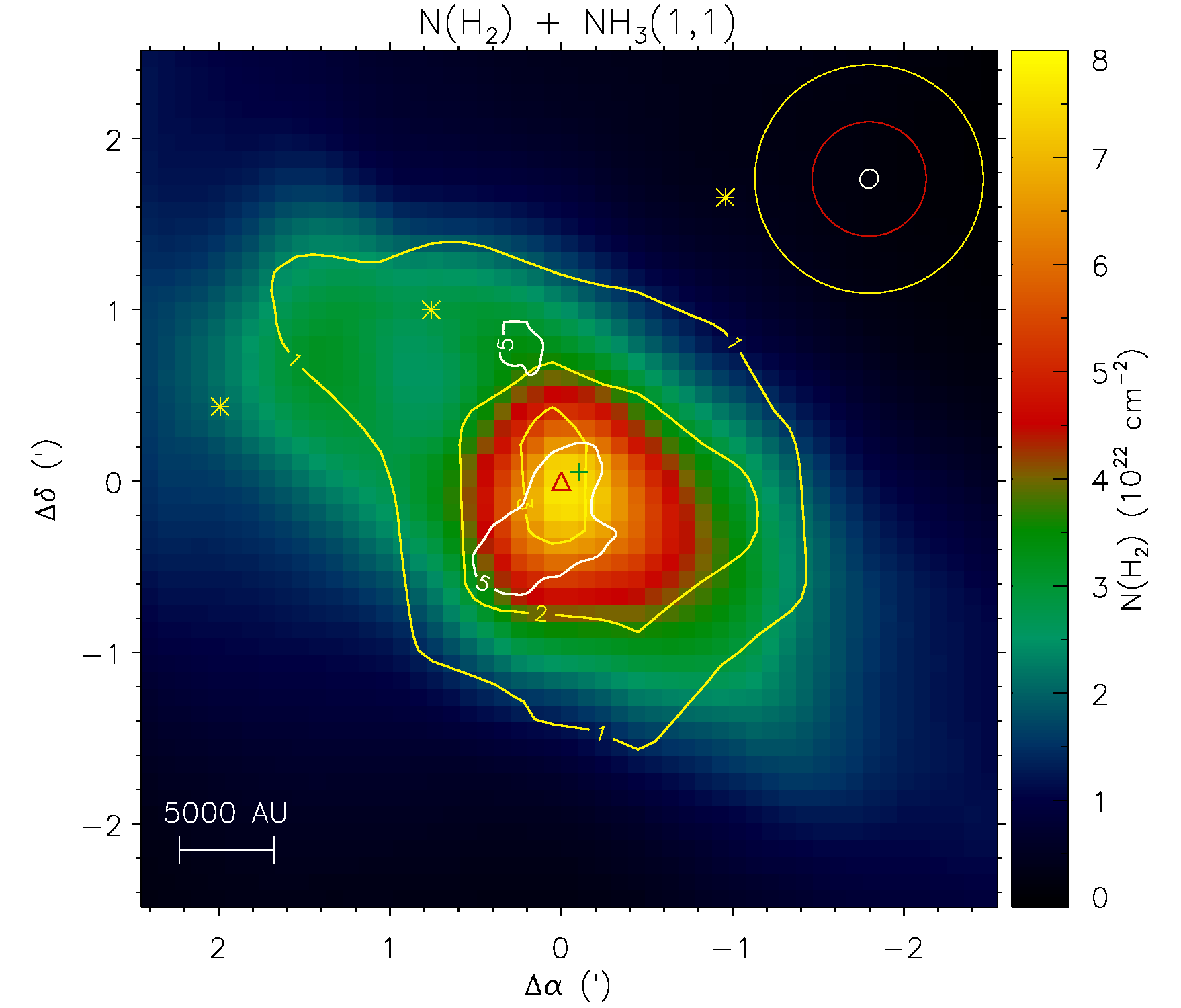} 
\end{picture}}
\put(60,0){\begin{picture}(0,0)
\includegraphics[width=6.5cm,angle=0]{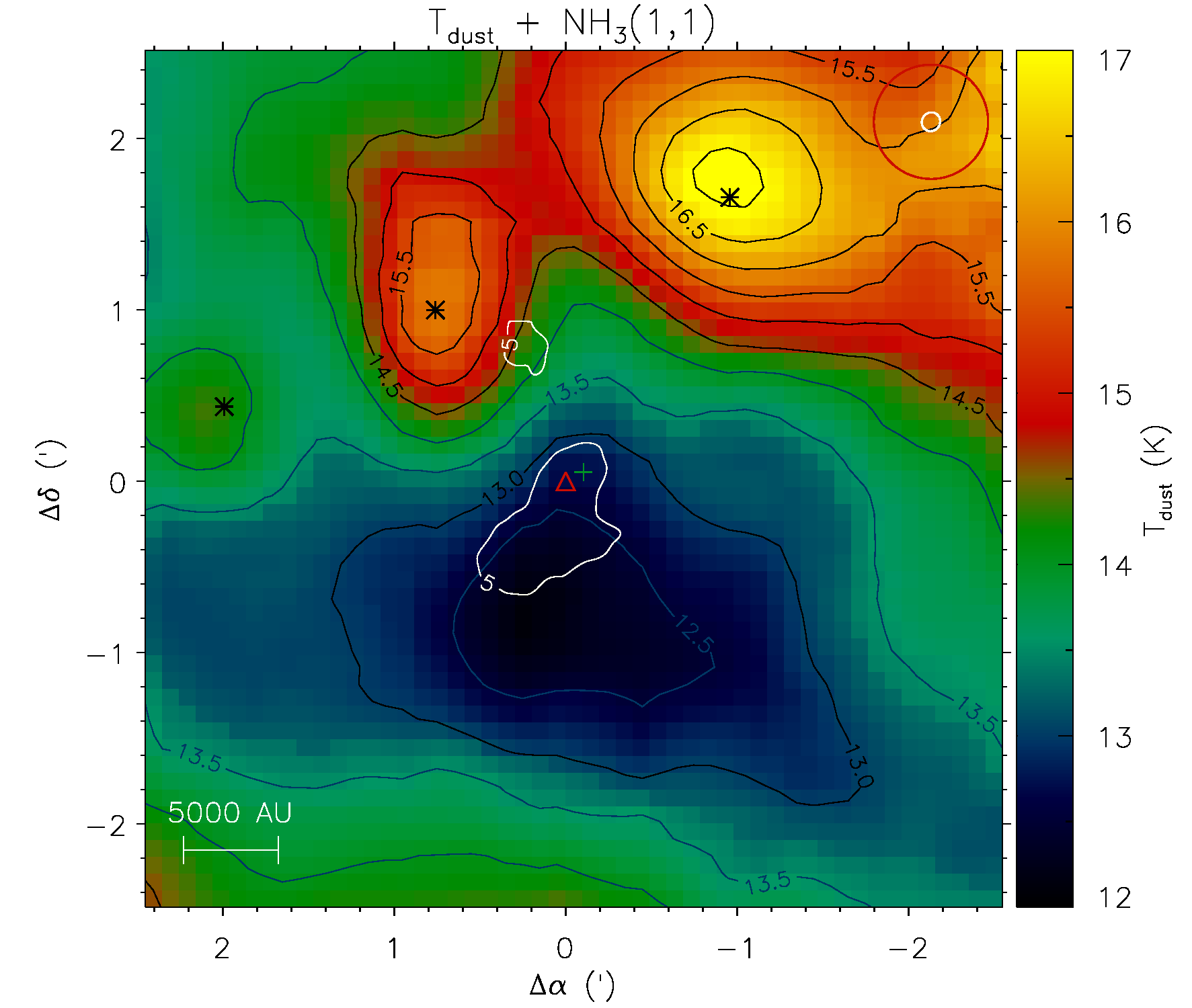} 
\end{picture}}

\end{picture}
\caption{{\bf Left:} Intensity of thermal dust emission
  (MJy~sr$^{-1}$) at the wavelength $\lambda=250$ $\mu$m in a region
  of $5\arcmin\times5\arcmin$ region around Cha-MMS1. The map is
  centred at the \textit{Spitzer} 24-$\mu$m peak indicated with a red
  triangle. The red circle in the top right corresponds to the
    average \textit{Herschel} beamsize (FWHM, $18\arcsec$) at 250
    $\mu$m. The small white ellipse shows the synthesised ATCA beam
    ($\sim7\arcsec$). {\bf Middle:} Dust temperature, $T_{\rm dust}$
  (K), in the same region. The red circle in the top right
    represents the resolution of the calculated $T_{\rm dust}$ and
    H$_2$ column density maps ($40\arcsec$). {\bf Right:} H$_2$
  column density, $N({\rm H_2})$ ($10^{22}$ cm$^{-2}$) derived from
  the 250-$\mu$m optical thickness, $\tau_{250\,\mu{\rm m}}$. The dust
  temperature and column density maps are derived by fitting a
  modified blackbody function to the \textit{Herschel} intensity maps
  at 500, 350, 250, and 160 $\mu$m assuming an emissivity index
    of $\beta=2.0$ (see text). The locations of three prominent young
  stellar objects (YSOs), Ced110 IRS2 (right), IRS4 (middle), and IRS6
  (left), are indicated with asterisks. The plus sign indicates the
  location of the millimetre source Cha-MMS1a (\citealt
  {1996A&A...314..258R}).  The white contour show the integrated
  intensity level 5 K km s$^{-1}$ of the NH$_3(1,1)$ satellites as
  observed with ATCA. The yellow contours in the right show the
  integrated intensity ($\int T_{\rm A}{\rm dv}$) of the NH$_3(1,1)$
  line emission (K km s$^{-1}$) observed at Parkes. The Parkes
    beam is shown with the large yellow circle in the top right.
}
\label{figure:Herschel250}
\end{figure*}

We used far-infrared and submm maps from the \textit{Herschel}
satellite to inspect the overall distribution of gas and dust in the
vicinity of Cha-MMS1.  The \textit{Herschel} maps used in this study were
extracted from the extensive mapping of the Chamaeleon I (Cha I) cloud
with the Spectral and Photometric Imaging Receiver (SPIRE) and
Photodetector Array Camera and Spectrometer (PACS) instruments, as
part of the {\sl Herschel Gould Belt Survey}
(\citealt{2010A&A...518L.102A}; \citealt{2012A&A...545A.145W}; see
also {\tt http://www.herschel.fr/cea/gouldbelt/en/}). An overview of
the \textit{Herschel Space Observatory} is given in
\citet{2010A&A...518L...1P}.The SPIRE and PACS instruments are
described in \citet{2010A&A...518L...3G} and
\citet{2010A&A...518L...2P}, respectively.

The pipeline-reduced data of Cha I are publicly available at the
Herschel Science Archive\footnote{\tt
  http://herschel.esac.esa.int/Science\_Archive.shtml}.  The
wavelengths covered in the SPIRE+PACS parallel mode survey are 500,
350, and 250 $\mu$m (SPIRE), and 160 and 70 $\mu$m (PACS). The FWHM
values of the gaussians fitted to the beam profiles are, in order of
decreasing wavelength, 37$\arcsec$, 25$\arcsec$, 18$\arcsec$,
12$\arcsec\times$16$\arcsec$, and 6$\arcsec\times$12$\arcsec$, the
PACS beams being clearly elongated in observations carried out in the
parallel mode. We did not apply any zero-point corrections to the
  pipeline-reduced maps.

In order to derive the distributions of the dust temperature, $T_{\rm
  dust}$, and the optical thickness,, $\tau_{\lambda}$, of the dust
emission, the maps at 500, 350, 250, and 160 $\mu$m were convolved
with Gaussians to a common resolution of $40\arcsec$ (FWHM), and the
intensity distributions were fitted pixel by pixel with a modified
blackbody function, $I_\nu \approx B_\nu(T_{\rm dust})\tau_\nu \propto
B_\nu(T_{\rm dust}) \, \nu^{\beta}$, where $B_\nu(T)$ is the
  Planck function, and $\tau_\nu$ is the optical thickness of the
  emitting source at the frequency $\nu$. The modified blackbody
function characterises optically thin ($\tau_\nu \ll 1$) thermal
dust emission at far-IR and submm wavelengths.  We fixed the
emissivity/opacity exponent to $\beta=2.0$.  The observationally
derived values of $\beta$ are typically close to 2.0 (e.g.,
\citealt{2010ApJ...708..127S,2011A&A...527A.111J}), although higher
values have been reported for cold regions
(e.g. \citealt{2011A&A...536A..23P}). An average value of $\sim 1.8$
was found in a recent study of the Taurus-Auriga molecular cloud, but
with significant anticorrelation between dust temperature and the
spectral index \citep{2011A&A...536A..25P}.

The pipeline-calibrated in-beam flux densities were colour corrected to
monochromatic flux densities at the standard wavelengths in the course
of $T_{\rm dust}$ fitting using the adopted shape of the source spectrum
(a modified blackbody spectrum with $\beta=2.0$) and the spectral
response functions of SPIRE and PACS photometers for an extended
source.\footnote{see the SPIRE and PACS Observer's Manuals (Versions 2.4) at 
{\tt http://herschel.esac.esa.int}, ''Herschel Documentation''}   
In addition, the SPIRE pipeline flux densities were converted from point
source calibration to extended source calibration before the
temperature fitting and colour correction.

The errors of  $T_{\rm dust}$ and $\tau_{\rm 250\mu m}$ caused by
  photometric noise were estimated by a Monte Carlo method using
  the $1\sigma$ error maps provided for the three SPIRE bands and a
  7\% uncertainty of the absolute calibration for all four bands
  according to the information given in SPIRE and PACS manuals (SPIRE
  Observers' Manual, Version 2.4; PACS Observer's Manual, Version 2.4)
  Different realisations of the $T_{\rm dust}$ and $\tau_{\rm 250\mu m}$
  maps were calculated by combining the four intensity maps with the
  corresponding error maps, assuming that the error in each pixel is
  normally distributed. The errors in each pixel were obtained from
  the standard deviation of one thousand realizations. The relative
  errors of $T_{\rm dust}$ obtained this way are typically 2--3 \%, and the 
  errors for $\tau_{\rm 250\mu m}$ are
  around 10\%.
    
The 250-$\mu$m optical thickness map, calculated from
 $\tau_{\rm 250\mu m}= I_{\rm 250\,\mu m}/B_{250\mu{\rm m}}(T_{\rm
dust})$, is proportional to the molecular hydrogen column density,
$N({\rm H_2})$. The column density can be obtained from
\begin{equation}
N({\rm H_2}) = \frac{\tau_{250\mu{\rm m}}}{\kappa_{250\mu{\rm m}} {\bar m}_{\rm H_2}},
\end{equation}
where $\kappa$ is the dust ``opacity'', or absorption cross-section
per unit mass of {\sl gas}, and ${\bar m}_{\rm H_2}$ ($=2.8$ amu) is
the average particle mass {\sl per} H$_2$ {\sl molecule} (assuming
10\% He and negligible amout of metals).  For the dust opacity at $250\,\mu$m we adopt
$\kappa_{250\mu{\rm m}}^{\rm g} = 0.10\,{\rm cm^{2}\, g^{-1}}$ 
\citep{1983QJRAS..24..267H}.

The column density map in a $5\arcmin\times5\arcmin$ region around
Cha-MMS1 is shown in the rightmost panel of
Fig.~\ref{figure:Herschel250}.  The other panels of
Fig.~\ref{figure:Herschel250} show the $250\,\mu$m intensity map
($I_{250\,\mu{\rm m}}$, left) and the $T_{\rm dust}$ map (middle).
  The YSOs Ced 110 IRS6 (Class I), IRS4 (Class I), and IRS2 (Class
  III) are surrounded by warm dust and they are visible as local maxima
  on the $T_{\rm dust}$ map.  The cavity on the western side of IRS2
  is partly visible on the $T_{\rm dust}$ map [see Fig.~4 of
  \citet{2006AJ....132.1923B}]. Cha-MMS1 is situated in a cold region
south of IRS2. IRS4 has been found to produce an outflow which
probably collides into the Cha-MMS1 core \citep{2007ApJ...664..964H,
  2011ApJ...743..108L}.

To estimate the effect of the uncertainty related to the adopted
  emissivity index, $\beta$, we repeated the analysis described above
  using fixed emissivity indices of $\beta=1.8$ and $\beta=2.2$.
  Lowering $\beta$ leads to higher dust colour temperatures, and lower
  optical thicknesses $\tau_{\rm 250\,\mu m}$. The effect of increasing $\beta$
  is the opposite. The maximum ${\rm H_2}$ column density is found
  slightly southeast of Cha-MMS1a at the offset
  $(+4\arcsec,-8\arcsec)$, and the colour temperature minimum lies
  further southeast at the offset $(+23\arcsec,-50\arcsec)$. The
  range of $T_{\rm dust}$ and $N({\rm H_2})$ values at these two positions for
  $\beta=1.8-2.2$ are listed in Table~\ref{table:Td+NH2_vs_beta}.

\begin{table}
  \caption{$T_{\rm dust}$ and $N({\rm H_2})$ values at the temperature minimum 
and the column density maximum using the emissivity index $\beta=2.0$, and 
the range $\beta=1.8-2.2$.} 
 
\centering
\label{table:Td+NH2_vs_beta}
\small
\begin{tabular}{l l l}\hline\hline
$T_{\rm dust}$ minimum & R.A. $11^{\rm h}06^{\rm h}38\fs6$  & 
                   Dec. $-77\degr24\arcmin22\arcsec$ \\
                 &  $T_{\rm dust}$ (K) &  $N({\rm H_2})$ ($10^{22}$ cm$^{-2}$) \\
$\beta=2.0$      &  $12.1\pm0.3$       &    $3.8\pm0.4$                   \\
$\beta=1.8-2.2$      &  $12.7-11.5$       &   $3.0-4.9$  \\ \hline

$N({\rm H_2}))$ maximum & R.A. $11^{\rm h}06^{\rm h}32\fs8$  & 
                   Dec. $-77\degr23\arcmin40\arcsec$ \\
                 &  $T_{\rm dust}$ (K) &  $N({\rm H_2})$ ($10^{22}$ cm$^{-2}$) \\
$\beta=2.0$      &  $12.6\pm0.3$       & $7.7\pm0.8$   \\
$\beta=1.8-2.2$      &  $13.3-12.0$       & $6.0-9.8$   \\

\end{tabular}
\end{table}

The uncertainties owing to photometric noise are about 3\% and 10\%
for $T_{\rm dust}$ and $N({\rm H_2})$, respectively. The change of
$\beta$ by 0.2 has an effect which corresponds to about two times the
photometric noise. The line-of-sight inhomogeneities, especially in dense clouds
cause further uncertainties as discussed in e.g. \cite{2012A&A...547A..11N}, 
\cite{2012A&A...542A..21Y}, and \cite{2013A&A...555A.140S}.

\section{Results}

\label{results}

\subsection{Parkes results}

\label{Parkes_results}

The general concept of deriving gas parameters from the hyperfine
  spectra is presented by \cite{1979ApJ...234..912H}. The
  spectroscopic properties of ammonia can be found, e.g., in the
  papers of \cite{1967PhRv..156...83K}, \cite{1975ApJS...29...87P} and
  in the review by \cite{1983ARA&A..21..239H}. The standard method of
  deriving ammonia column densities from the NH$_3$ inversion lines is
  described in, e.g., \cite{1986A&A...157..207U} (their Appendix).  We
  have used the CURVEFIT function of IDL 
  (Interactive Data Language\footnote{Distributed by Exelis Visual Information Solutions, Inc. (http://www.exelisvis.com)}) 
  to fit 18 Gaussians to the hyperfine components of the ${\rm
    NH_3(1,1)}$ inversion line. The parameters of this fit are the
  excitation temperature ${T_{\rm ex}}$, total optical thickness
  $\tau_{\rm tot}(1,1)$ (the sum of the peak optical thicknesses of
  the 18 components), line-of-sight velocity $V_{\rm LSR}$ and
  linewidth ${\Delta V}$.  As typical for cold cores, the $(2,2)$
  hyperfine satellite groups are very weak also in Cha-MMS1, and the
  optical thickness of this line cannot be estimated. The $(2,2)$
  column density is derived assuming the main group is optically thin
  and that ${T_{\rm ex}} (2,2)={T_{\rm ex}}(1,1)$.

A hyperfine component fit to the Parkes NH$_3(1,1)$ spectrum at the
offset $(10\arcsec,-14\arcsec)$ shown in
Fig.~\ref{figure:Parkes_spectrum} gives the following parameters:
$T_{\rm MB,peak}=1.9\pm0.2$ K (the main-beam brightness temperature
assuming the efficiency $\eta_{\rm MB}=0.7$), $V_{\rm
  LSR}=4.42\pm0.02 \, \kmps$, $\Delta V = 0.48\pm0.03 \, \kmps$
(FWHM), $\tau_{\rm tot} = 8.8\pm1.1$.  The standard method of deriving
the kinetic temperature from the $(1,1)$ to $(2,2)$ comparison gives
$T_{\rm kin} = 11.1\pm1.2$ K. Assuming (somewhat unrealistically)
uniform beam filling we obtain an excitation temperature of $T_{\rm
  ex}(1,1) = 4.8\pm0.2$ K, and a column density of $N(p{\rm
    NH_3}) = 2.7\pm0.4 \times10^{14}$ cm$^{-2}$ for {\sl para}-${\rm
    NH_3}$. In cold clouds $N(p{\rm NH_3}) \approx N({\rm
    NH_3(1,1)})$ as other metastable para states, $(2,2)$, $(4,4)$,

  etc. are scarcely populated. The obtained column density represents
  the average {\sl para}-${\rm NH_3}$ column density within the $80\arcsec$
  Parkes beam. The corresponding H$_2$ column density at
$(10\arcsec,-14\arcsec)$, as derived from the \textit{Herschel}
observations and smoothed to the same angular resolution, is $N({\rm
  H_2})=3.3\times10^{22}$ cm$^{-2}$. The implied fractional
  {\sl para}-ammonia abundance is $X(p{\rm NH_3}) \sim 1\times10^{-8}$.

\subsection{ATCA wide-band images}

\label{widebandimages}

\begin{figure}
   \centering
   \includegraphics[width=5.8cm,angle=0]{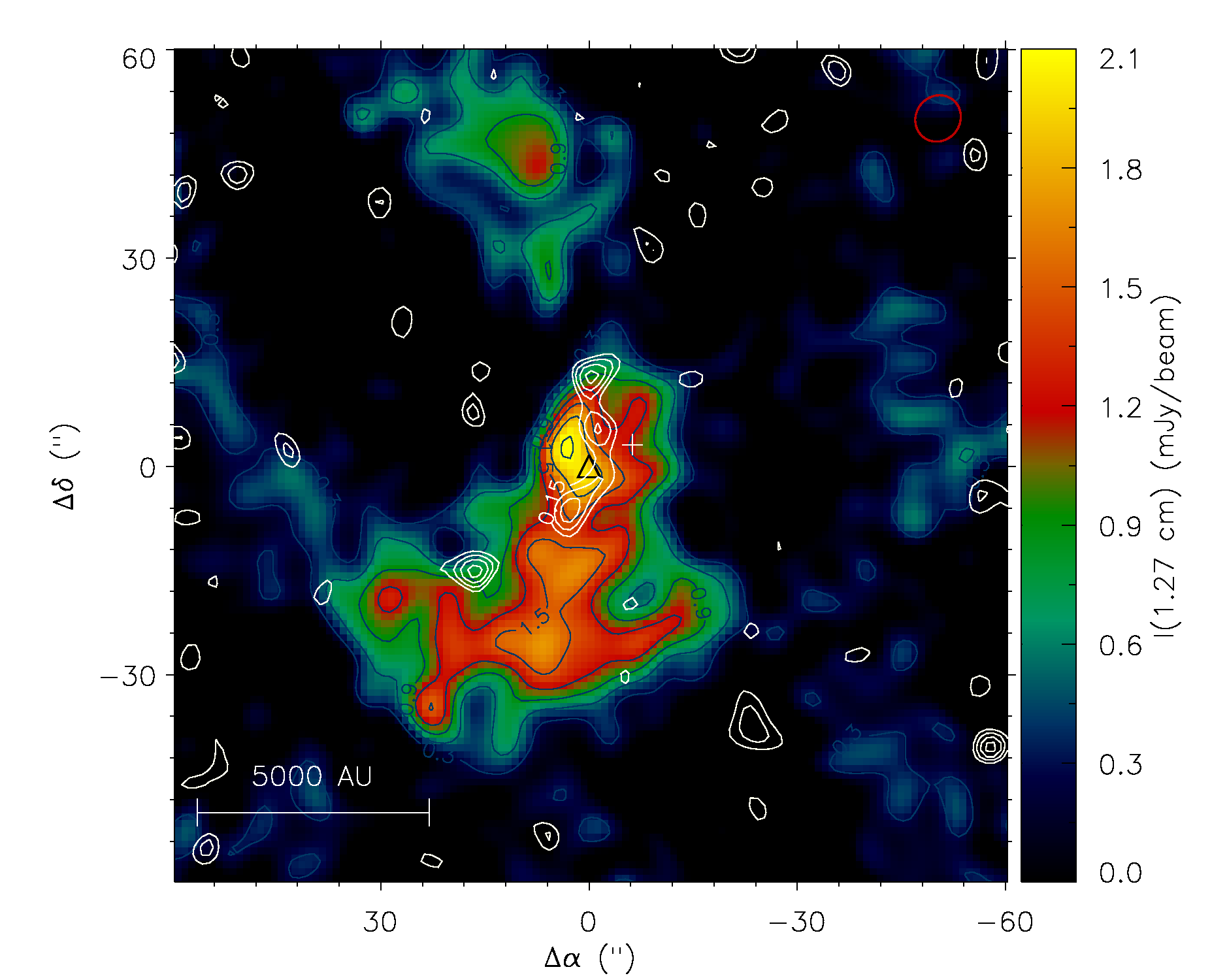}
   \includegraphics[width=5.8cm,angle=0]{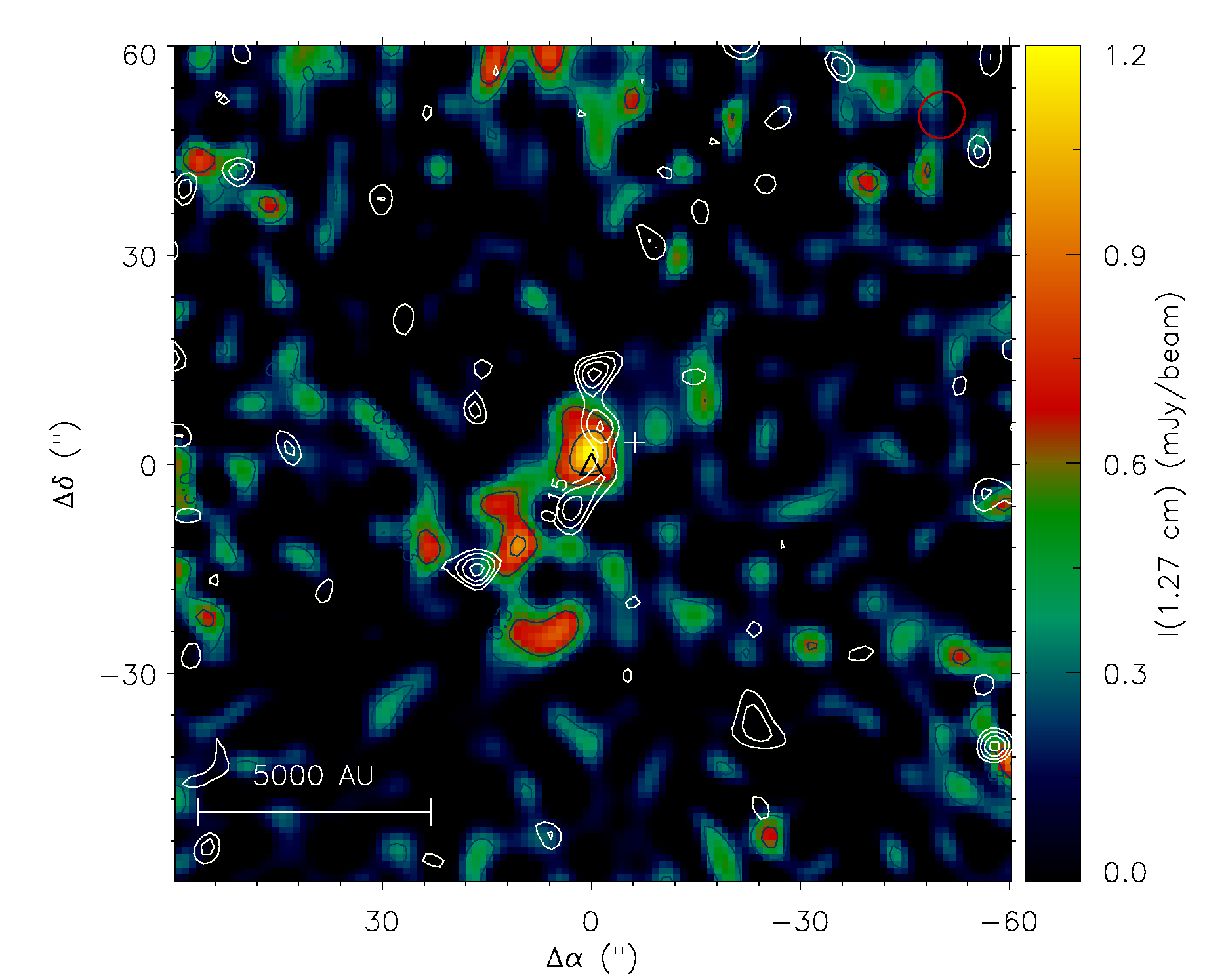}

   \caption{ATCA wide-band images.  {\bf Top:} The pixel image shows
     the intensity distribution of the two central channels covering
     the ${\rm NH_3}(1,1)$ line. {\bf Bottom:} The pixel image of
     the channel containing the $(2,2)$ line. The contour map
     superposed on both images represents the true 1.3 cm continuum
     emission, i.e. wide-band channels without ammonia line
     contamination. The positions of Cha-MMS1a and the
     \textit{Spitzer} 24-$\mu$m maximum are marked with a plus sign
     and a triangle, respectively.}
    \label{figure:widebandimages}
\end{figure}

The 112 MHz band covers the sky frequencies of the NH$_3(1,1)$ and
$(2,2)$ lines. The $(1,1)$ hyperfine groups are split between the two
central channels (7 and 8), centred at 23\,688.5 and 23\,692.5 MHz, and
one of the other channels (centred at 23\,720.5 MHz) can have
contribution from the $(2,2)$ line with the rest frequency 23\,722.6
MHz. Three separate images have been formed from the
wide-band visibility data using a) the two channels
with $(1,1)$ lines; b) the channel with the (2,2); and c) the remaining
11 line-free channels. The first two (a and b) are shown as pixel images in
Fig.~\ref{figure:widebandimages}, and the continuum map (c) is superposed as contours 
on these.

The intensity in the two central channels is clearly higher than in
other channels, and the distribution is similar to that of the
integrated satellite intensity shown in Fig.~\ref{figure:Satellite_area_map}.
This indicates that emission in channels 7 and 8 comes primarily from
the NH$_3(1,1)$ line. The channel with the $(2,2)$ line shows faint
compact features located in the region with strong $(1,1)$ emission. The
distribution is different from that of the pure con\-ti\-nuum shown as
contours. Also the channel covering the $(2,2)$ line frequency is likely
to be dominated by ammonia.

Three emission peaks at a level of 1 mJy~beam$^{-1}$ (corresponding to
about 5 K km~s$^{-1}$) can be identified on the $(2,2)$ map.
The offsets of the $(2,2)$ peaks with respect to the Cha-MMS1a
position, and estimates of the kinetic temperature, $T_{\rm kin}$, in
these locations are presented in Table~\ref{table:Tkin}. Two of the
$(2,2)$ peaks ({\bf a} and {\bf c} in the Table) coincide with the $(1,1)$
peaks in the bright ridge, while the third peak ({\bf b}) lies about
$10\arcsec$ east of the line connecting the $(1,1)$ peaks.  Position
({\bf a}) is the brightest in both $(1,1)$ and $(2,2)$ emission, and
its coordinates are R.A. $11^{\rm h}06^{\rm m}33\fs5$,
Dec. $-77\degr23\arcmin33\arcsec$ (J2000). This position lies
  $4\arcsec$ west of the \textit{Herschel} peak given in Table 1 of
  \cite{2012A&A...545A.145W}, and $2\arcsec$ north of the
  \textit{Spitzer} 24-$\mu$m maximum.

\begin{table}
\caption{Integrated intensities of the NH$_3(2,2)$ lines and $(1,1)$
satellites and kinetic temperatures derived from these towards three
$(2,2)$ peaks. The integrated $(2,2)$ intensities are estimated from the
wide-band channel covering the sky frequency of the line.}  \centering
\label{table:Tkin}
\
\begin{tabular}{crrccc} \hline\hline
Pos. & $\Delta\alpha$ & $\Delta\delta$ & $\int T_{\rm B}(2,2) {\rm d}\upsilon$ & 
 $\int T_{\rm B}(1,1,s) {\rm d}\upsilon$ &  $T_{\rm kin}$ \\
 & ($\arcsec$)&($\arcsec$)& (K~km~s$^{-1}$) & (K~km~s$^{-1}$) & (K) \\ \hline
{\bf a} & 6 &  -1&  $6.2\pm0.4$ & $13.7\pm0.2$ & $20\pm2$ \\
{\bf b} &17 & -15&  $4.7\pm0.4$ & $9.2\pm0.2$  & $22\pm3$  \\
{\bf c} &11 & -27&  $4.5\pm0.4$ & $11.1\pm0.2$ & $19\pm3$ \\ 
\hline
\end{tabular}
\end{table}

The temperature estimates have been derived from the $(1,1)$ and
$(2,2)$ intensities following the method first presented by
\citet{1979ApJ...234..912H} assuming local thermodynamic equilibrium
(LTE) and optically thin emission.  The relation derived by
\citet{1983A&A...122..164W} and \citet{1988MNRAS.235..229D} have been
used to convert the rotational temperature, $T_{12}$, to the kinetic
temperature, $T_{\rm kin}$.

From the $(1,1)$ linewidths one can obtain the following upper limits
for the kinetic temperatures at the positions listed in
Table~\ref{table:Tkin} {\bf a}: 33 K ($\Delta \upsilon = 0.30\, {\rm
  km\,s}^{-1}$), {\bf b}: 23 K ($\Delta \upsilon = 0.25 \, {\rm
  km\,s}^{-1}$), and {\bf c}: 20 K ($\Delta \upsilon = 0.23 \, {\rm
  km\,s}^{-1}$). These results are consistent with the estimates from
the integrated $(1,1)$ and $(2,2)$ intensities suggesting that
the maxima of the $(2,2)$ map pin-point warm spots in the core.

\vspace{2mm}

The pure continuum map shows a string of weak sources near the
ammonia maximum. The intensities of the four peaks near the map centre
are only 3-4 times the rms noise in the region, i.e. in the range
$0.26-0.33$ mJy~beam$^{-1}$. The tentative detection remains therefore
to be confirmed, or disproved, by a deeper integration.
\cite{2003A&A...401.1017L} derived a 3-$\sigma$ upper limit of 0.1
mJy/$4\arcsec$ beam for the intensity of Cha-MMS1 at 3.5 cm. The
confirmation of the 1.3 cm maximum intensity quoted above would mean a
spectral index of $\alpha > 1.2$ which would be consistent with
optically thick free-free emission.

\subsection{ATCA spectral line image}

\label{ATCA_spectral_images}

\begin{figure*}[ht]
   \centering
   \includegraphics[width=9.0cm,angle=0]{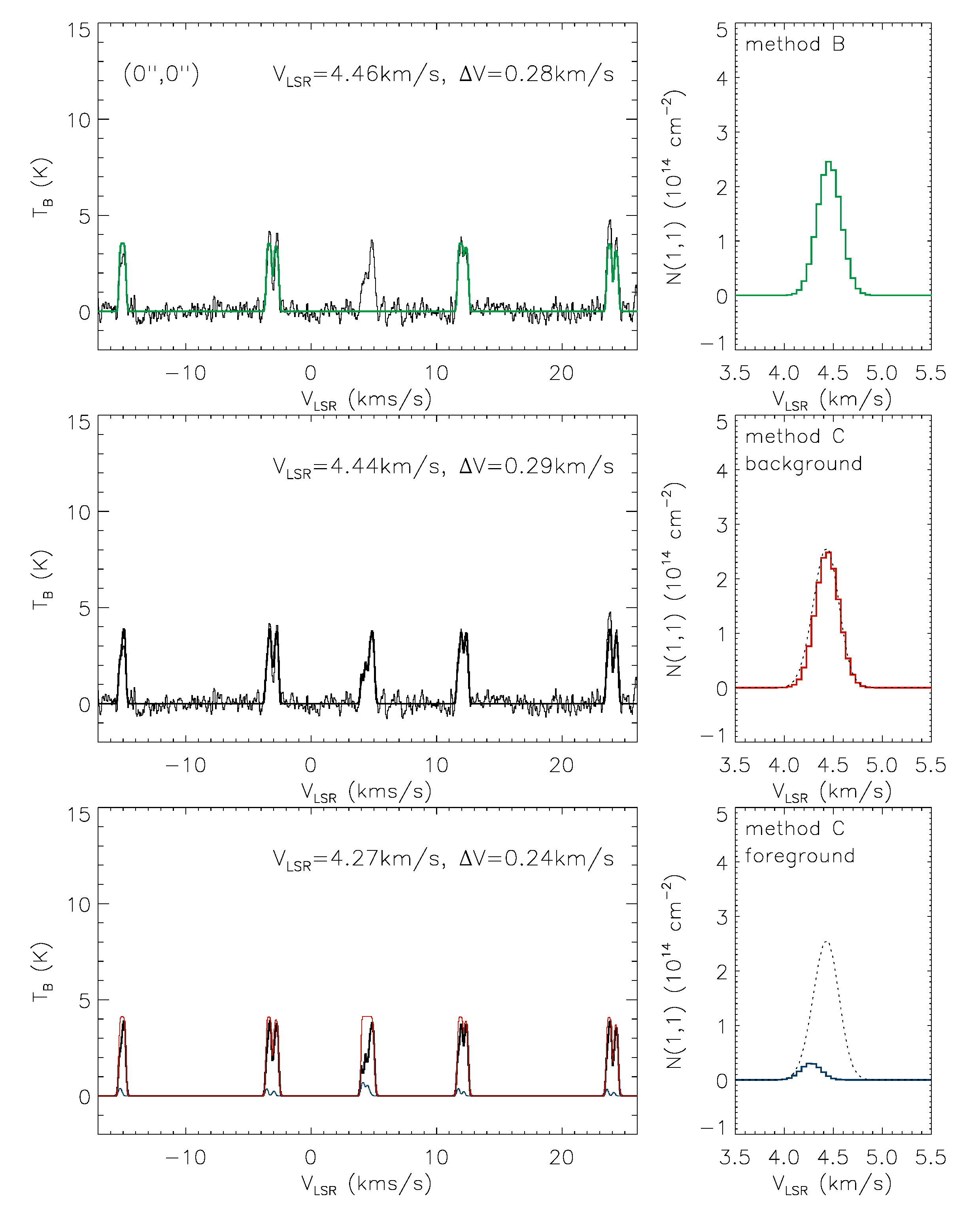}
   \includegraphics[width=9.0cm,angle=0]{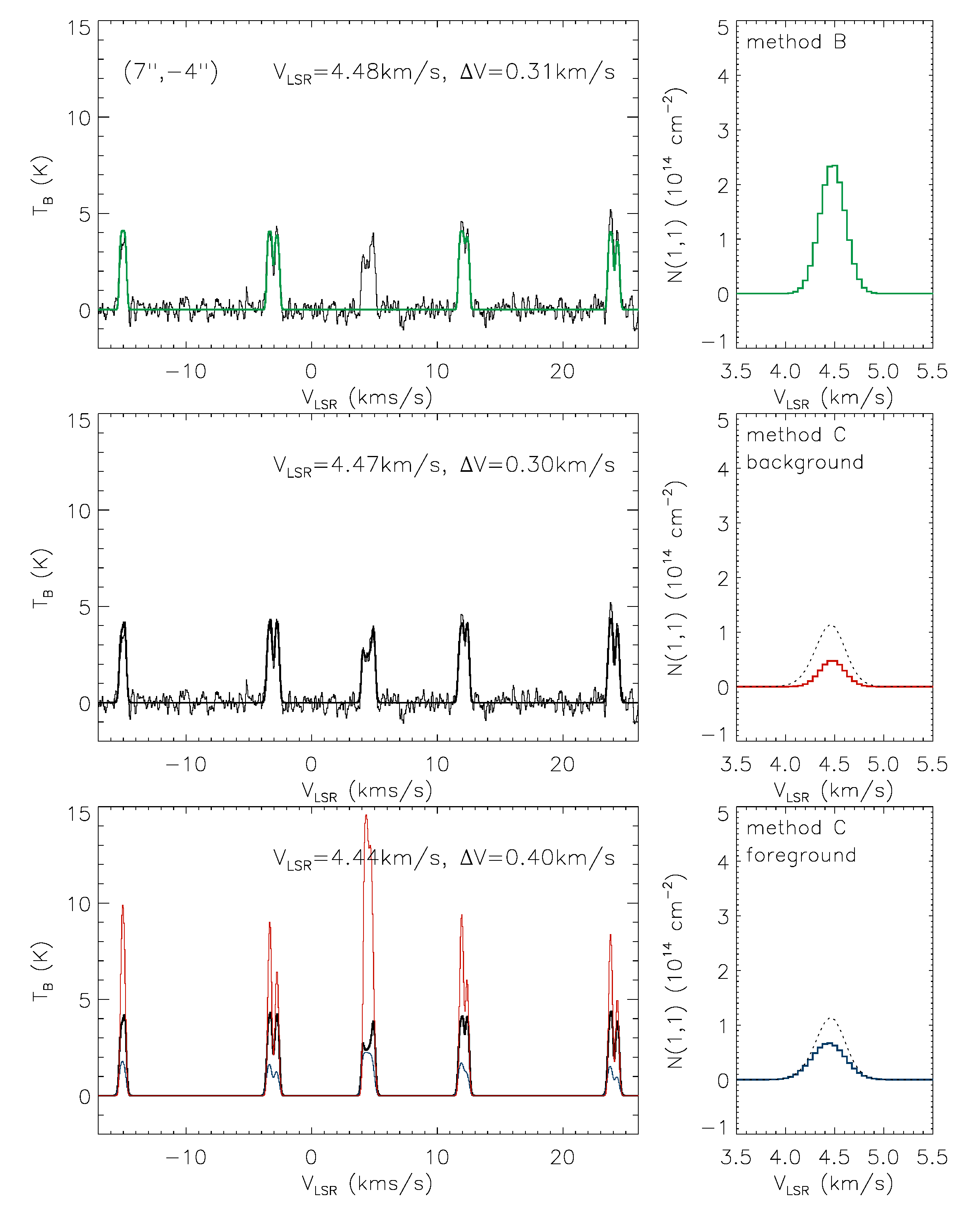}
   \caption{ATCA NH$_3(1,1)$ spectra towards the offsets $(0\arcsec,0\arcsec)$ and
     $(+7\arcsec,-4\arcsec)$ from Cha-MMS1a. The latter position
     coincides with a \textit{Spitzer} 24-$\mu$m emission peak.  
      The top panels show a Gaussian
     fit to the four satellite groups of the hyperfine structure
     (green curve). The column density spectrum of ammonia in the
     rotational level $(1,1)$ derived from these fits are shown on the
     right side of the satellite spectra. The panels in the middle row show 
     two-component fits to the full hyperfine structure including the main group 
     (black curve). In the bottom panels, the modelled emission profiles from 
     the two layers (component 1: red, component 2: blue) are shown separately.
     The foreground component 2 has a lower excitation temperature than
     component 1, and causes absorption features. The corresponding NH$_3(1,1)$ 
     column density spectra of components 1 (red) and 2 (blue) are shown on the 
     right. The dashed curve shows the combined NH$_3(1,1)$ column density spectrum of layers 1 and 2.}
\label{figure:ATCA_spectra}
\end{figure*}

The ATCA narrow-band visibility data from both observing runs were
combined and inverted into an $xy$v-image cube.  The original
  spectral visibility data contain 1\,024 channels. After dropping 82
  channels at both ends of the bandpass the final image cube has 860
  velocity (frequency) channels 0.05 \kmps (3.9 kHz) in width. The
frequency range covers all the five hyperfine groups of the
$(J,K)=(1,1)$ inversion line. Two NH$_3(1,1)$ spectra extracted
  from the ATCA image cube are shown in
  Fig.~\ref{figure:ATCA_spectra}.  The spectra are Hanning smoothed
  averages over $3\arcsec\times3\arcsec$ regions around the indicated
  positions. The selected positions are $(0\arcsec,0\arcsec)$
  (Cha-MMS1a) and $(7\arcsec,-4\arcsec)$ (a local maximum in the
  24-$\mu$m \textit{Spitzer} map).

In the optically thin case the integrated intensity of the main group
in the centre of the spectrum (including the transitions $F_1=|{\bar
  J}+{\bar I}_{\rm N}|=1\rightarrow1$, $F_1=2\rightarrow2$, where
$|{\bar I}_{\rm N}|=1$ is the nuclear spin of the N nucleus) is equal
to the total integrated intensity of the four satellites (from left to
right: $F_1=1\rightarrow0$, $F_1=1\rightarrow2$, $F_1=2\rightarrow1$,
$F_1=0\rightarrow1$). The groups are further split owing to magnetic
interactions.

Near the NH$_3(1,1)$ maximum the main hyperfine group is sometimes
weaker than any of the four satellites. This indicates
self-absorption.  On the outskirts of the map the spectra look 
  optically thin, the main group being the strongest.  The $(1,1)$
satellites are less affected by large column densities than the main
group because the total optical thickness of each satellite is about
one quarter of that of the main group. In the optically thick limit
the five groups have equal intensities.

  We have tested three methods of modelling the $(1,1)$ hyperfine
  structure. The method selected depends on the ratio of the 
  integrated main group intensity to the integrated intensity of the 
  satellites. We denote this ratio by M/S. 

  A) In the first method, Gaussian fits to all the 18 hyperfine
  components were made in regions where the optical thickness does not
  appear to be very large. The method was applied in the positions
  where ${\rm M/S}\ge 1/2$. This requirement is fulfilled at the
  core edges.

  B) In the regions where ${\rm M/S}<1/2$, we have made Gaussian fits
  only to the ten hyperfine components of the four satellite
  groups. For these values of M/S, the optical thickness of the main
  group is large, and satellite line structure is better reproduced
  using this model than using method A. 
  
  C) In the third method, the spectra with ${\rm M/S}<1/4$ (suggesting
  self-absorption) have been fitted with a function describing
  emission from two contacting plane-parallel homogeneous layers at
  different excitation temperatures, $T_{\rm ex,fg}$ and $T_{\rm
    ex,bg}$. Here we indicate quantities pertinent to the component in
  the foreground with ``fg'', and those related to the background with
  ``bg''.  The two-layer method is discussed in
  \cite{1997A&A...326..366A}. Assuming that LTE holds in both layers, and that the 
   radiation impinging on the background layer is the cosmic microwave background,
   the brightness temperature spectrum can be written as
\begin{equation}
\begin{array}{lcl}
T_{\rm B}(\upsilon) &=& T_{11} \, [F(T_{\rm ex,bg})-F(T_{\rm CMB})]
              (1-e^{-\tau_{\rm bg}(\upsilon)})e^{-\tau_{\rm fg}(\upsilon)} + \\
             & & T_{11} \, [F(T_{\rm ex,fg})-F(T_{\rm CMB})]
               (1-e^{-\tau_{\rm fg}(\upsilon)})  \; ,
\end{array}
\end{equation}
where $T_{11} \equiv h\nu_{11}/k$, $\nu_{11}$ is the line frequency, $h$ and $k$ are the 
Planck and Boltzmann constants, $T_{\rm CMB}$ is the cosmic background temperature, and 
the function $F(T)$ is defined by
$$
F(T) \equiv \frac{1}{e^{T_{11}/T}-1} \; .
$$ 
When the background component has a higher excitation temperature than
the foreground component, i.e., when $T_{\rm ex,bg} > T_{\rm ex,fg}$ the
spectra show absorption dips.

\begin{center}
  \begin{table*}
   \caption{NH$_3(1,1)$ line parameters towards Cha-MMS1a, 
   $(0\arcsec,0\arcsec)$, the \textit{Spitzer} 24-$\mu$m peak, $(7\arcsec,-4\arcsec)$,
   using the methods A, B, and C explained in the text.}
   
 \begin{tabular}{llllll} \hline\hline
$(0\arcsec,0\arcsec)$ & $T_{\rm ex}$ (K) & $\tau_{\rm tot}$  & $\upsilon_{\rm LSR}$ 
(km\,s$^{-1}$) & $\Delta \upsilon$ (km\,s$^{-1}$) & 
$N(1,1)$ ($10^{15}\,{\rm cm}^{-2}$) \\ \hline
A    & $5.81\pm0.05$ &$101\pm17$   &  $4.458\pm0.003$ & $0.265\pm0.008$ & $2.2\pm0.4$ \\
B    & $6.31\pm0.08$ &$60\pm8$     &  $4.456\pm0.003$ & $0.284\pm0.009$ & $1.5\pm0.2$ \\
C bg  & $6.89\pm0.19$ &$61\pm8$     &  $4.437\pm0.004$ & $0.286\pm0.009$ & $1.5\pm0.2$ \\
C fg  & $3.53\pm0.29$ &$7.4\pm1.4$  &  $4.271\pm0.012$ & $0.244\pm0.023$ & $0.16\pm0.06$ \\ 
\hline
$(7\arcsec,-4\arcsec)$ &  $T_{\rm ex}$ (K)& $\tau_{\rm tot}$ & $\upsilon_{\rm LSR}$ 
(km\,s$^{-1}$) & $\Delta \upsilon$ (km\,s$^{-1}$) & $N(1,1)$ ($10^{15}\,{\rm cm}^{-2}$) 
\\ \hline
A   & $6.34\pm0.05$ &$80\pm9$ &  $4.471\pm0.003$ & $0.296\pm0.008$ & $2.1\pm0.2$ \\
B   & $6.87\pm0.10$ &$53\pm5$      &  $4.478\pm0.003$ & $0.310\pm0.008$ &  $1.5\pm0.2$ \\
C bg & $18.0\pm1.9$  &$10.8\pm1.4$  &  $4.474\pm0.003$ & $0.298\pm0.007$ & $0.31\pm0.05$ \\
C fg & $5.01\pm0.13$ &$15.0\pm1.2$  &  $4.437\pm0.008$ & $0.402\pm0.016$ & $0.57\pm0.06$ \\ \hline
\end{tabular}
  \label{table:fit_parameters}
  \end{table*} 
\end{center}

  The use of methods B and C are demonstrated in
  Fig.~\ref{figure:ATCA_spectra} which show NH$_3(1,1)$ spectra
  towards the offsets $(0\arcsec,0\arcsec)$ and $(7\arcsec,-4\arcsec)$, together
  with Gaussian fits to the hyperfine stucture. The fit to the
  satellites only (method B) is shown as a green curve in the top
  panel. The components of the two-layer fit (method C) are shown as
  red and blue curves in the bottom panel, and the modelled total
  emission spectrum is shown as a black curve superposed on the
  observed spectra in the middle panel.  The NH$_3(1,1)$ column
  density spectra (i.e., the column density of ammonia in the
  rotational level $(1,1)$ in each velocity channel) derived from the
  fits are shown on the right of Fig.~\ref{figure:ATCA_spectra}. The
  total $(1,1)$ column density can be obtained by summing the values
  in each channel. The NH$_3(1,1)$ line parameters for the spectra
  shown in Fig.~\ref{figure:ATCA_spectra}, derived using all the three
  methods are listed in Table~\ref{table:fit_parameters}. In this
  table we list the the excitation temperature, $T_{\rm ex}$, of the
  $(1,1)$ transition (assumed to be the same for all components), the
  sum of the peak optical thicknesses of the 18 hyperfine
  components, $\tau_{\rm tot}$, the LSR velocity of the line, $\upsilon_{\rm LSR}$,
  the linewidth (the FWHM of an individual hyperfine component),
  $\Delta \upsilon$, and the NH$_3$ column density in the rotational state $(1,1)$, 
  $N(1,1)$. The symbol ``C bg'' means fit to the background component 
  in method ``C'', and ``C fg'' means fit to the foreground
  component.

\begin{figure}[htb]
   \centering
   \includegraphics[width=8.0cm,angle=0]{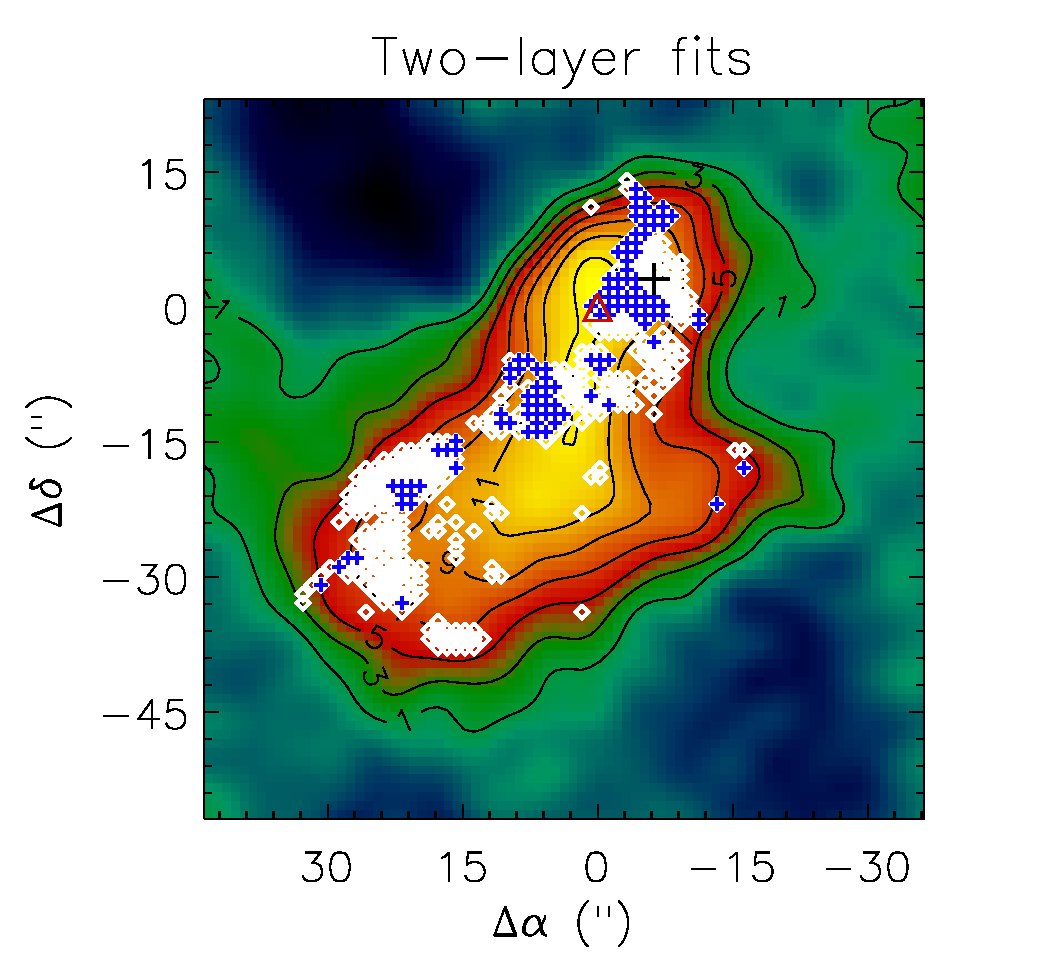}
   \includegraphics[width=7.0cm,angle=0]{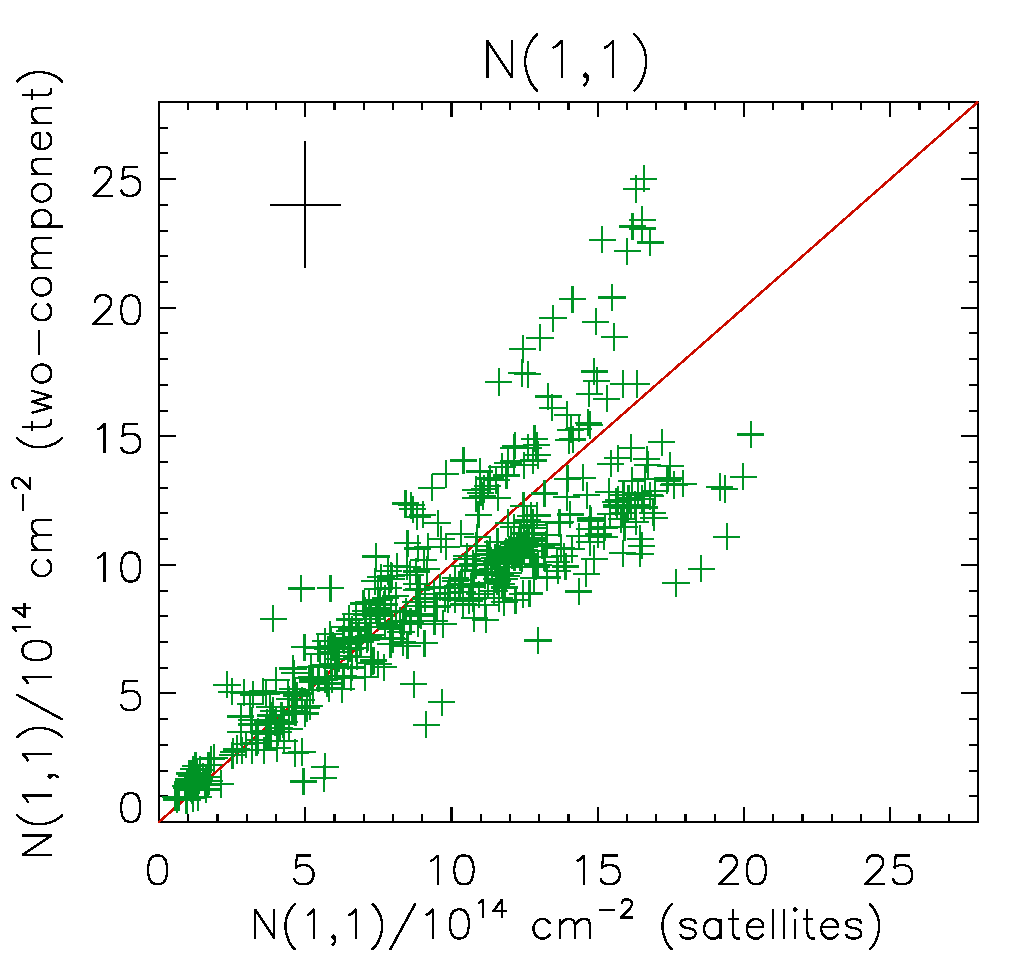}
   \caption{{\bf Left:} The region where the two-layer fit (method C)
     is applicaple is shown with white diamonds on the integrated
     brightness temperature map of the NH$_3(1,1)$ satellites. Blue
     plus signs indicate positions where $T_{\rm ex,bg} > 11$ K, i.e.,
     where the excitation temperature of the background component
     exceeds the average kinetic temperature of the core. {\bf Right:}
     Comparison between the NH$_3(1,1)$ column densities derived by
     fits to the hyperfine satellites only (method B, $x$-axis) and 
     two-layer fits (method C, $y$-axis). The average uncertainties of the 
     points are indicated in the top left. The red line represents the 
     one-to-one correlation.}
\label{figure:two-layers}
\end{figure}

 The region where the two-layer fit gives reasonable values,
    i.e., where the four fit parameters ($T_{\rm ex}$, $\tau_{\rm
      tot}$, $\upsilon_{\rm LSR}$, and $\Delta \upsilon$) are physically meaningful
    (positive) for both layers, is marked in
    Figure~\ref{figure:two-layers}.  The region concentrates around a
    NW-SE axis going through the \textit{Spitzer} 24-$\mu$m peak and
    millimetre peak Cha-MMS1a. As discussed in Sect.~\ref{Velocity_distribution} this
    axis can be identified as the rotation axis of the dense nucleus
    of the core. The absorption occurs at velocities between $4.3$ and
    $4.5$ km\,s$^{-1}$,  around the systemic velocity of the
    cloud. Strong absorption is not seen on the northeastern nor the
    southwestern side of the axis because there the inner parts are
    either blue- or redshifted with respect to the absorbing envelope
    (see Sect.~\ref{Velocity_distribution}).

  The ammonia spectrum towards the position of the \textit{Spitzer}
  24-$\mu$m emission maximum, $(7\arcsec,-4\arcsec)$, shown in
  Fig.~\ref{figure:ATCA_spectra} (right panel) has one of the deepest
  main group absorption features in the whole map. The two-layer fit
  (see Table~\ref{table:fit_parameters}) suggests that the underlying
  component has a clearly higher $T_{\rm ex}$ than the foreground
  component. Probably also the kinetic temperature of component 1 is
  elevated with respect to average value of the core, $T_{\rm kin}
  \sim 11$ K, derived from Parkes data (Sect.~\ref{Parkes_results}),
  because for thermal emission lines $T_{\rm ex} \leq T_{\rm kin}$. It
  is evident from Fig.~\ref{figure:ATCA_spectra} (top panel) that the
  total column density from the fit to the satellites (method B, with
  $T_{\rm ex} \sim 6.9$ K) is substantially higher than that resulting
  from the two-layer fit. On the other hand, towards the position
  Cha-MMS1a, $(0\arcsec,0\arcsec)$, lying somewhat off from the brightest emission,
  the satellite fit corresponds closely to background component of the
  two-layer fit, and the total column densities from methods B and C
  are similar. The average excitation temperatures of the background 
  and foreground components are $\langle T_{\rm ex,bg}\rangle=10.5$ K and 
  $\langle T_{\rm ex,fg}\rangle=4.5$ K, respectively. The positions where the 
  $T_{\rm ex,bg} > 11$ K (the average gas kinetic temperature) are marked in
  Fig.~\ref{figure:two-layers} with blue plus signs.

\begin{figure}[hbt]
   \centering
   \includegraphics[width=9.0cm,angle=0]{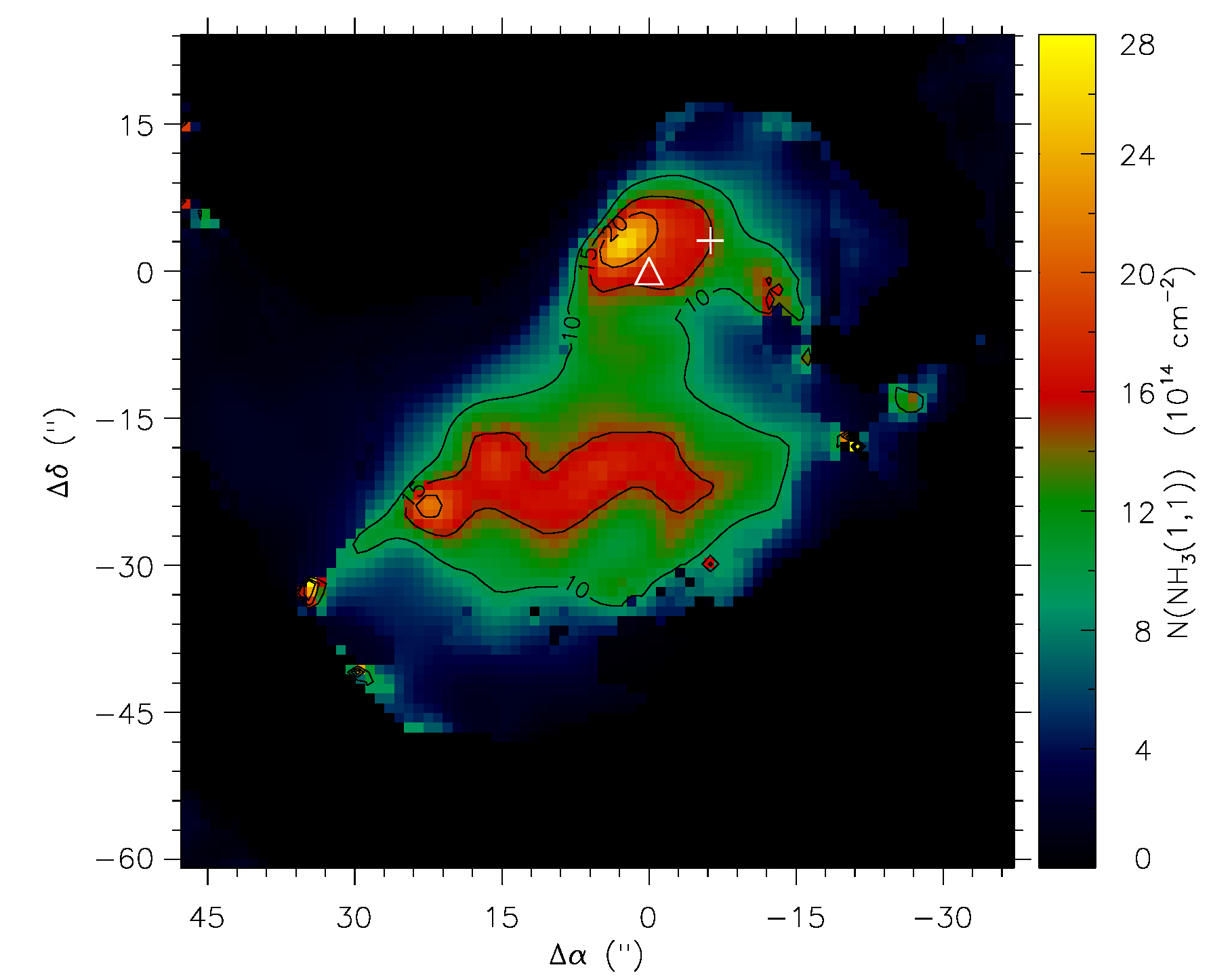}
   \caption{The column density of NH$_3$ in the rotational level
     $(J,K)=(1,1)$ in Cha-MMS1. The contour values are $10$, $15$, and
     $20\times10^{14}$ cm$^{-2}$. The positions of Cha-MMS1a and the
     \textit{Spitzer} 24-$\mu$m maximum are marked with a plus sign and
     a triangle, respectively.}
\label{figure:N11map}
\end{figure}

\begin{figure}[htb]
\begin{picture}(190,190)
\put(0,0){\begin{picture}(0,0)
\includegraphics[width=8cm,angle=0]{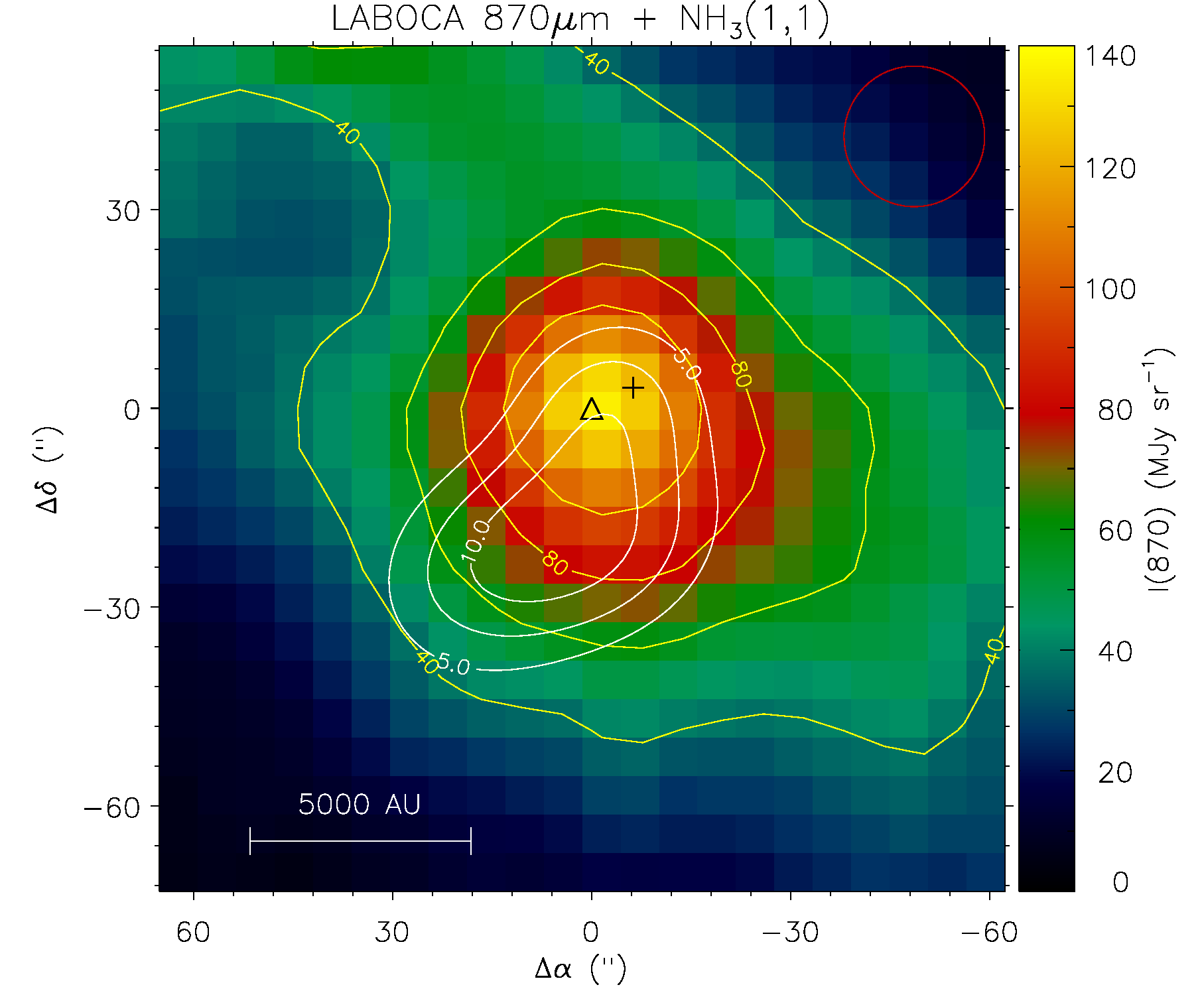} 
\end{picture}}
\end{picture}
\caption{The column density map of the NH$_3(1,1)$ smoothed to an
  angular resolution of $21\arcsec$ (contours) overlaid on the
    $870\, \mu$m map obtained with LABOCA (\citealt{2011A&A...527A.145B};
    \citealt{2013A&A...557A..98T}).  The contour levels are 5, 7.5, and
  $10\times10^{14}\, {\rm cm}^{-2}$. The plus sign indicates the
  1.2-mm dust emission peak Cha-MMS1a and the triangle shows the
    location of the \textit{Spitzer} 24-$\mu$m source.}
\label{figure:laboca}
\end{figure}

A comparison between the NH$_3(1,1)$ column densities derived by
  two-layer fits (method C), and fits to the hyperfine satellites only
  (method B) is shown in the right panel of
  Fig.~\ref{figure:two-layers}. One can see that the two column
  densities agree reasonably well up to $N(1,1) \sim 10^{15}$
  cm$^{-2}$, but towards the highest column densities discrepancies
  occur. Similar effects are seen in comparisons between methods A and
  B and between methods A and C. Method A tends to overestimate the
  optical thickness of the line (and the column density) in the region
  where the main component shows self-absorption.  On the other hand,
  the results from method C (altogether eight parameters for the two
  components) are uncertain when the radial velocities of the
  background and foreground components are nearly the same because
  both components have narrow linewidths. The average uncertainty of
  the two-layer fits is larger than for the satellite fits (see
  Fig.~\ref{figure:two-layers}). Owing to the mentioned uncertainties,
  we have used a combination of methods A and B, that is, satellite
  fits in the central region and fits to all the 18 hyperfine
  components in the outer parts, in the derivation of the NH$_3(1,1)$
  column density map. This map is shown in Fig.~\ref{figure:N11map}.
The column density map resulting from a combination of models A, B,
and C (the satellite fit replaced by the two-layer fit in the region
where the fit parameters are positive) is similar to that shown in
Fig.~\ref{figure:N11map}, except for a larger number of ``bad'' pixels
deviating from the overall pattern. A comparison between the
  NH$_3(1,1)$ column density map and the $870\,\mu$m dust emission map
  is shown in Fig.~\ref{figure:laboca}. The dust continuum map was
  observed by \cite{2011A&A...527A.145B} with the LABOCA bolometer
  array on APEX. In this figure, the ammonia map is smoothed to the
  angular resolution of the LABOCA map, $21\arcsec$.

The bright integral-shaped ridge (P.A. $175\degr$) visible in
the integrated intensity map (Fig.~\ref{figure:Satellite_area_map})
does not stand out in the column density map.  The highest column
densities are found on the NE side of the \textit{Spitzer} 24-$\mu$m
peak, and in a wiggled, filamentary structure in the southern part of
the core. The offsets of the two most prominent maxima from Cha-MMS1a
are $(+9\arcsec,0\arcsec)$ and $(+29\arcsec,-27\arcsec)$. The column
densities of NH$_3(1,1)$ at these positions are
$(2.8\pm0.5)\times10^{15}$ cm$^{-2}$ and $(2.3\pm0.5)\times10^{15}$
cm$^{-2}$, respectively. Judging from comparison with the
\textit{Herschel} and LABOCA maps (Figs.~\ref{figure:Herschel250} and
\ref{figure:laboca}), the southern ammonia maximum is not associated
with an H$_2$ column density maximum, and it may therefore indicate a
local enhancement in the ammonia abundance.

Because high resolution $(2,2)$ spectra are not available, we have
not been able to derive properly the rotational temperature and the
total ammonia column density in each position. At low temperatures
the $(1,1)$ column density is roughly equal to that of {\sl para}-NH$_3$.  
In Sect.~\ref{Core_parameters}, we will use the {\sl para}-${\rm NH_3}$ 
abundance derived in Sect.~\ref{Parkes_results} to make a rough estimate 
of the mass and other physical parameters of the ammonia core.

\subsubsection{Velocity distribution}
\label{Velocity_distribution}
\begin{figure*}[htbp]
\begin{picture}(100,380)
\put(10,10){\begin{picture}(0,0)
\includegraphics[width=18cm,height=12cm,angle=0]{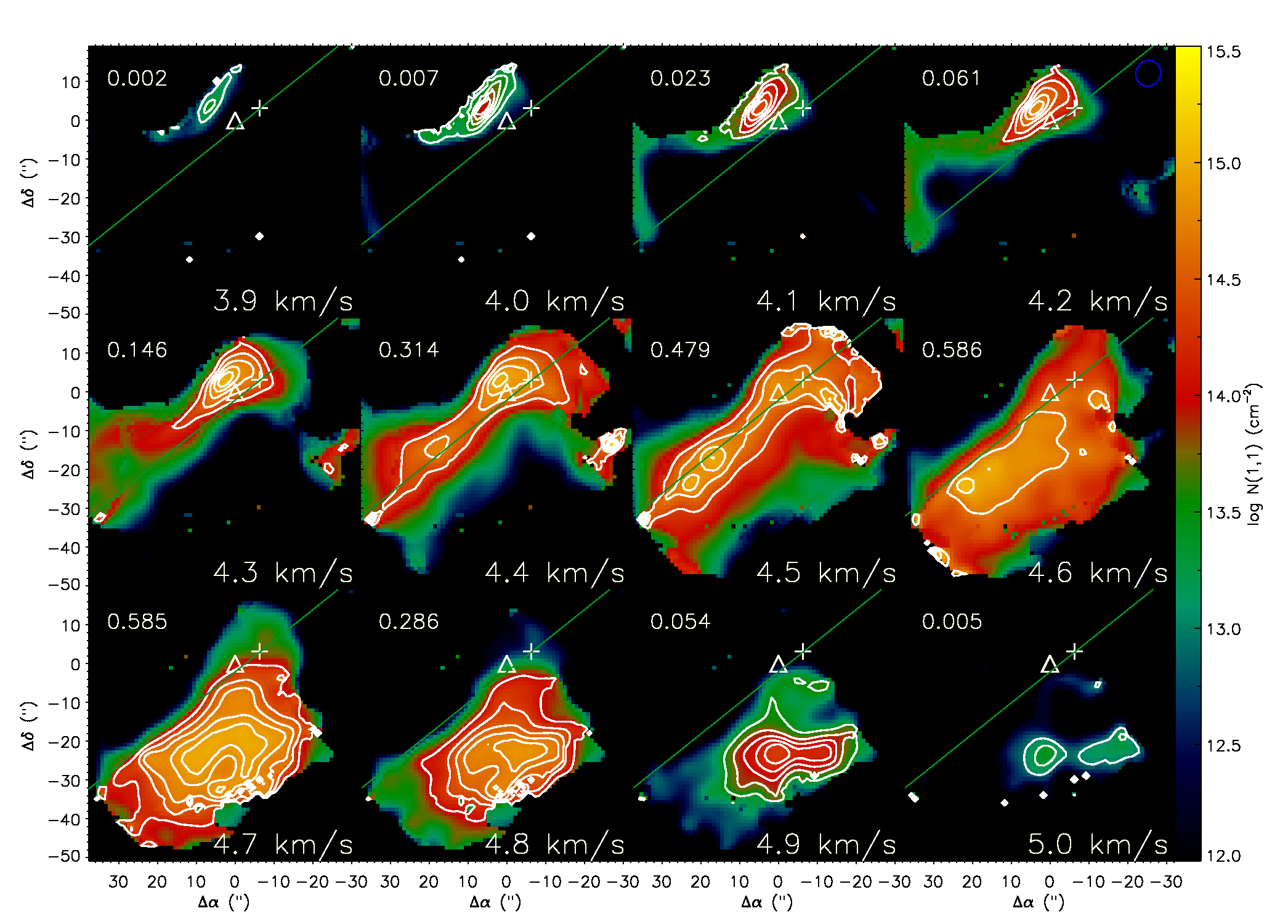} 
\end{picture}}
\end{picture}
\caption{Velocity channel maps of the NH$_3(1,1)$ column density
  in the Cha-MMS1 core. The column densities are shown using a logarithmic colour 
  scale. The LSR velocity in \kmps of the velocity channel is indicated in
  the bottom of each panel, and the mass in $M_\odot$ in this channel is indicated in the top left. The green line indicates the axis of rotation 
  determined from the velocity distribution. The positions of
  Cha-MMS1a and the \textit{Spitzer} 24-$\mu$m peak are marked with a plus sign and a 
 triangle, respectively.}
\label{figure:channelmap}
\end{figure*}

The NH$_3(1,1)$ column density spectra derived in the previous Section
were used to study the velocity structure of the core.  Velocity
channel maps formed from the column density cube are shown in
Fig.~\ref{figure:channelmap} as contours superposed on the
\textit{Spitzer} 24-$\mu$m dust emission map.  Each channel covers an
interval of 0.1 \kmps (two correlator channels). The centre velocities
range from 3.9 to 5.0 \kmps.  Channel maps calculated directly from
the spectral cube using the most blueshifted (leftmost) hyperfine
group ($F_1=1\rightarrow0$, with two close-lying components
$F=3/2\rightarrow 1/2$ and $F=1/2\rightarrow 1/2$) show the same
structure but with a higher noise level.

The core has a velocity gradient with the $\upsilon_{\rm LSR}$ increasing
from the NE to SW. Around the systemic velocity of the cloud, at
  4.4 and 4.5 \kmps one can see a linear structure with a position
  angle of $130\degr$.  The velocity gradient is thus almost planar
  near the NW-SE axis defined by this structure, and the axis can be
  identified as the rotation axis of the inner core. Gas components
  with velocities below 4.5 \kmps (i.e., blueshifted gas) can be found
  on the northeastern side of the axis, while the southwestern side
  has larger velocities (redshifted gas). The most blueshifted gas can
  be found on the NE side of the ammonia column density maximum and
  the 24-$\mu$m peak. The most redshifted gas lies some $25\arcsec$
  south of the \textit{Spitzer} position.

\begin{figure}[htbp]
  \centering
   \includegraphics[width=7.5cm,angle=0]{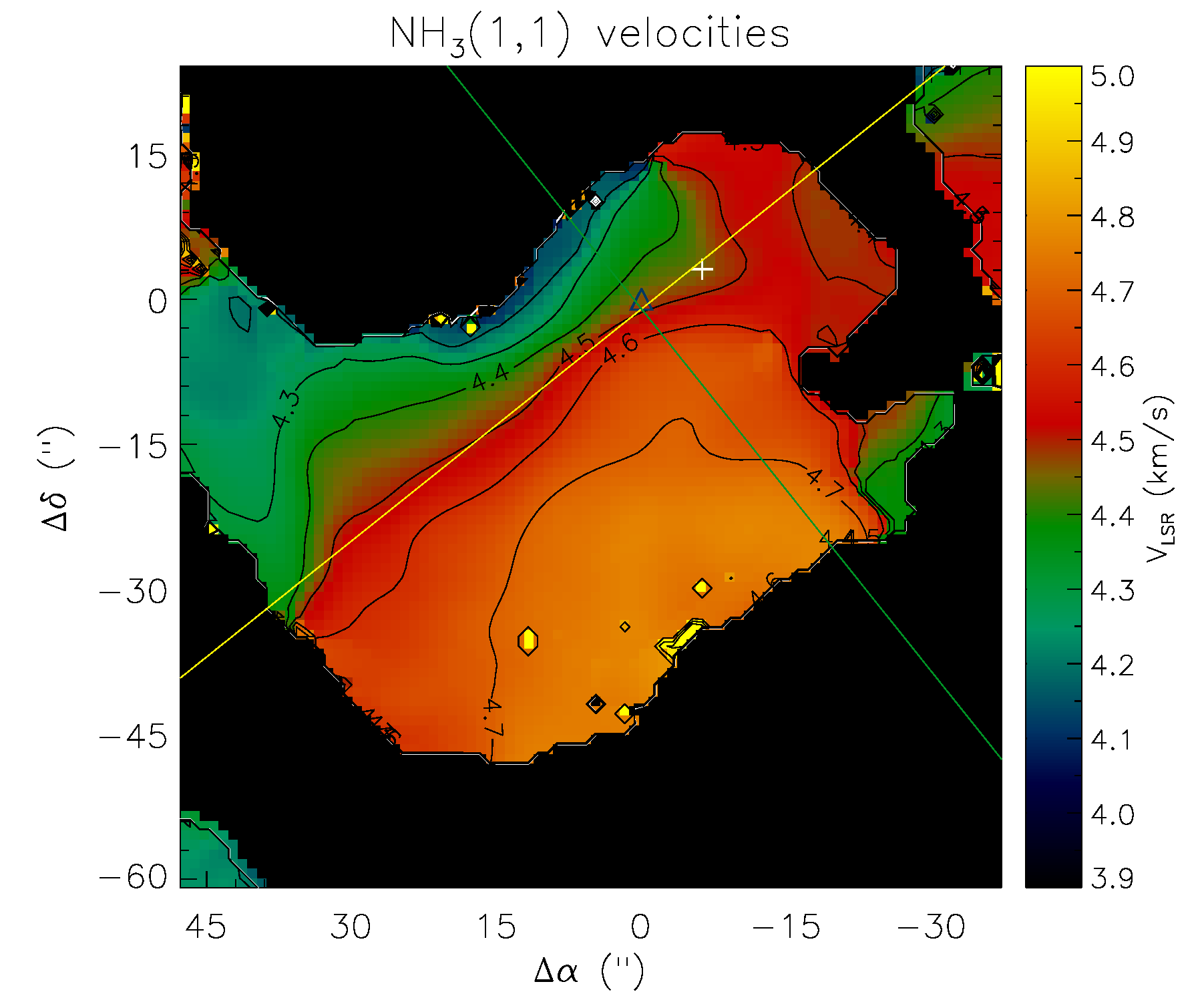}
   \includegraphics[width=7cm,angle=0]{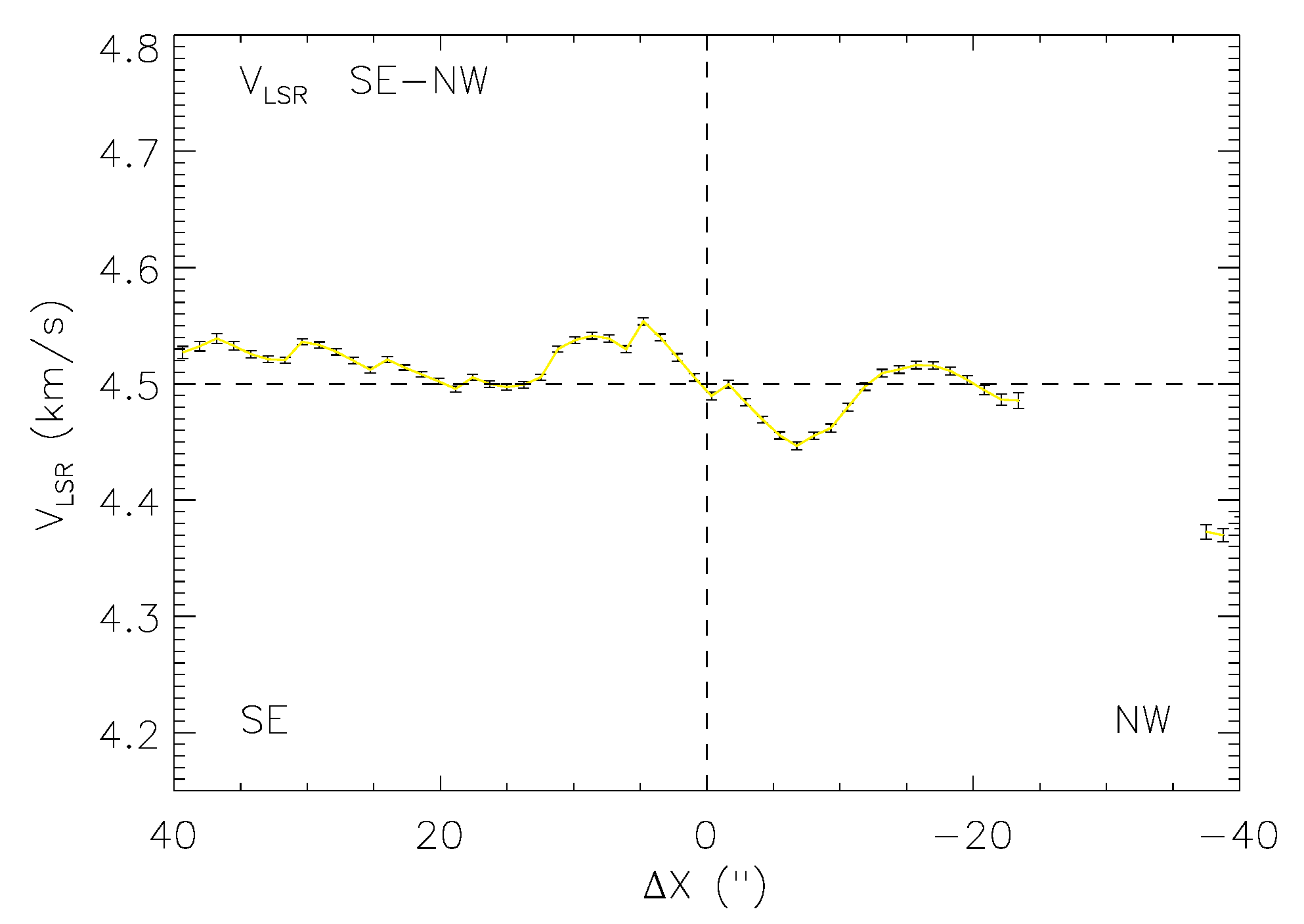}
   \includegraphics[width=7cm,angle=0]{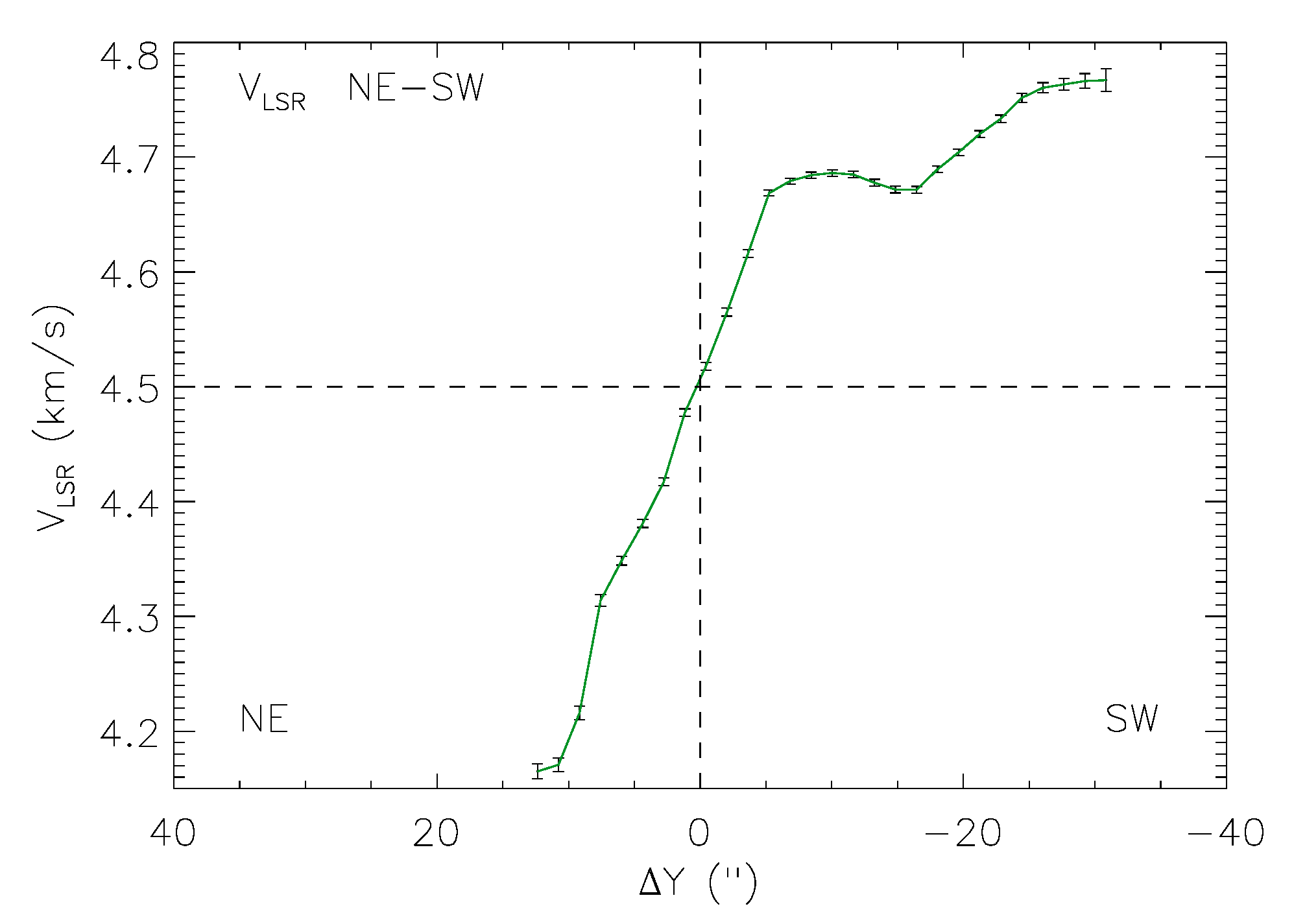}

   \caption{Distribution of the NH$_3(1,1)$ line velocities derived
     from Gaussian fits to the hyperfine structure. {\bf
       Top:} The radial velocity map. The $V_{\rm LSR}$ values can be
     read from the contours and the color bar on the right. The
     locations of the \textit{Spitzer} 24-$\mu$m source and the submm
     source Cha-MMS1a are shown with a triangle and a plus sign,
     respectively.  {\bf Middle:} The velocity profile along the NW-SE
     oriented axis drawn in yellow on the map, {\bf Bottom:} The
     velocity profile along the NE-SW oriented axis indicated with
     green. Also the error bars result from Gaussian fits to the $(1,1)$
     hyperfine structure.}
    \label{figure:vgradient}
\end{figure}

  The velocity gradient in the core is also illustrated in
  Fig.~\ref{figure:vgradient} which shows the radial velocity
  distribution as a pixel image, and position velocity diagrams
    along two perpendicular axes, one along the suggested rotation
  axis (P.A. $130\degr$) and another in the direction of the steepest
  velocity gradient (P.A. $40\degr$). The average velocity gradient
  along the second axis, estimated from the velocities at the
    ends of the axis, is about 20 \kmps pc$^{-1}$ (see
  Fig.~\ref{figure:vgradient}c). The two axes intersect at the
  \textit{Spitzer} 24-$\mu$m peak.

\begin{figure}[htbp]
  \centering
   \includegraphics[width=7.5cm,angle=0]{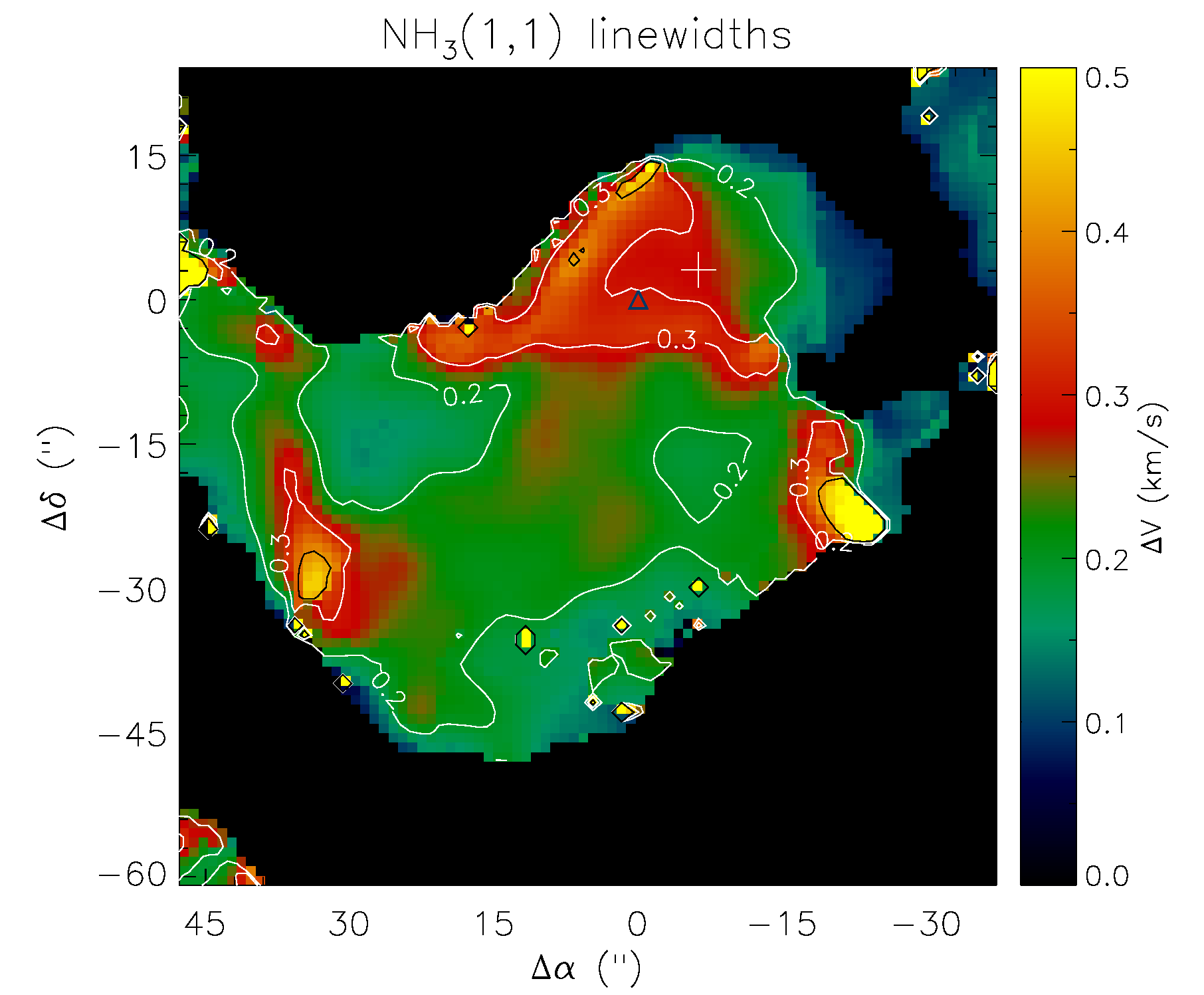}
    \caption{Map of the NH$_3(1,1)$ linewidths derived
     from Gaussian fits to the hyperfine structure.}
    \label{figure:widthmap}
\end{figure}

The distribution of the NH$_3(1,1)$ linewidths resulting from
Gaussian fits to the hyperfine structure is shown in
Fig.~\ref{figure:widthmap}. One can see that the lines are mostly
quite narrow, between 0.2 and 0.3 \kmps, but there is a slight increase in 
the linewidths ($\Delta V > 0.3$ \kmps) near the \textit{Spitzer} 24-$\mu$m source.

\subsection{Physical parameters of the core}

\label{Core_parameters}

 The column density of ammonia in the $(1,1)$ rotational
  level derived in Sect.~\ref{ATCA_spectral_images} have been used to
  estimate the mass and dynamical parameters of the core.  In the
  conversion from $N({\rm NH_3}(1,1))$ to $N({\rm H_2})$ needed for
  the mass, we have approximated the {\sl para}-${\rm NH_3}$ column
  density by the $(1,1)$ column density, and assumed a fractional 
  {\sl para}-ammonia abundance of $X(p{\rm NH_3}) = 1.0 \times10^{-8}$ derived
  from comparison between the Parkes and \textit{Herschel} data (see
  Sect. 3.1).

The results of the calculations are presented in
Table~\ref{table:parameters}. The parameters given in this Table refer
to the region above the column density threshold $N(1,1) > 5\times10^{14}$
cm$^{-2}$. The mass of this structure is $\sim 2\, M_\odot$. This mass
is slightly larger than the virial mass implied by the condition $E_{\rm grav} \,
+ \, 2 \, E_{\rm kin} = 0$, where $E_{\rm grav}$ is the gravitational
potential energy and $E_{\rm kin}$ is the kinetic energy including the
contributions of the thermal and turbulent motions and the core
rotation. The specific angular momentum of the core, $j = {\cal L}/M$,
given in the Table is calculated with respect to axis indicated in
Figs.~\ref{figure:channelmap} and \ref{figure:vgradient}
(P.A. $130\degr$). The obtained ratio of the rotational and
gravitational energies, $\beta_{\rm rot}\sim 0.07$, is close to the
high end of the range of values determined previously by
\cite{1993ApJ...406..528G}, \cite{1998ApJ...504..207B}, and
\cite{2002ApJ...572..238C} through single-dish observations.  The 
specific angular momentum is similar to that in the inner
part of L1544 as derived from interferometric NH$_2$D data by
\cite{2007A&A...470..221C}, and an order of magnitude lower than $j$
derived from NH$_3$ in the same target. According to results of
previous interferometric surveys the value we obtain in Cha-MMS1,
$j \sim 10^{-3} \,{\rm km\,s^{-1}\, pc}$, is typical for
protostellar envelopes and lies clearly below the specific angular
momenta derived for prestellar cores (\citealt{1999sf99.proc..129O};
\citealt{2007A&A...470..221C}). On the other hand, both $\beta$ and
$j$ are higher by about a factor of two than the median values
for cores formed in the simulation of \cite{2010ApJ...723..425D} of
magnetised, self-gravitating molecular clouds with decaying
turbulence.  According to the results of \cite{2010ApJ...723..425D}
angular momenta derived observationally from projected velocity maps
overestimate the true values by a factor of $\sim 10$ because of the
omission of fluctuations in the 3D velocity field.

\begin{table}
\caption{The dynamical parameters of the core derived from the ammonia
spectra (see text). The values represent the region inside the region where 
$N(1,1) > 5\times10^{14}$ cm$^{-2}$.}  
{\centering
\label{table:parameters}
\begin{tabular}{ll} \hline\hline
$\langle N(p{\rm NH_3}) \rangle$  &  $1.1 \times 10^{15} \, {\rm cm}^{-2}$ \\
Mass$^a$ &  2.2 $M_\odot$  \\
Radius   & $23\arcsec$, $3500$ AU \\
C.M. &   $\Delta\alpha=10\arcsec$, $\Delta\delta=-18\arcsec$ \\
$\langle n_{\rm H_2}\rangle$ &  $1.6\times10^{6}$ cm$^{-3}$ \\
$\langle {\rm v}_{\rm LSR}\rangle$ & 4.60 km~s$^{-1}$\\
$\langle \Delta v\rangle$ & 0.23 km~s$^{-1}$ \\
Velocity gradient & 20  km~s$^{-1}$\,pc$^{-1}$ \\
                  & ($\Omega \sim 6.5\times10^{-13}$ s$^{-1}$) \\ 
$E_{\rm therm}$ & $35\times10^{34}$ J \\
$E_{\rm turb}$  & $2\times10^{34}$ J \\
$E_{\rm rot}$  & $10 \times10^{34}$ J \\ 
$E_{\rm grav}$  & $-148 \times10^{34}$ J  \\ 
$|E_{\rm grav}|/2\,E_{\rm kin}$ & 1.6 \\
$E_{\rm rot}/|E_{\rm grav}|$, $\beta_{\rm rot}$ & 0.07 \\
Specific angular momentum$^b$, $j$ & 
$1.6\times10^{-3} \;{\rm km \,s^{-1}\, pc}$ \\
 & $= 3.9\times10^{20} \; {\rm cm^2\,s^{-1}}$ \\
\hline

\end{tabular}}

{\footnotesize
$^a$ The assumed fractional {\sl para}-ammonia abundance is 
$X(p{\rm NH_3})=1.0\times 10^{-8}$.} 

{\footnotesize
$^b$ The specific angular momentum is calculated with respect to the axis 
indicated  with a yellow line in Fig.~\ref{figure:vgradient}.} 

\end{table}

\subsection{Spectral energy distribution and evolutionary stage of
  Cha-MMS1}

\label{Results:Sed}

To characterise the emission arising from Cha-MMS1, we constructed its
spectral energy distribution (SED) by combining the \textit{Herschel}
and \textit{Spitzer} data. The \textit{Herschel} data were first
convolved to the resolution of the SPIRE/500-$\mu$m band ($37\arcsec$
or 0.027 pc at the source distance). The PACS and SPIRE flux densities
at 160, 250, 350, and 500 $\mu$m were then determined in an
  aperture with a radius of $25\arcsec$. The corresponding values are
  $7.7\pm0.9$, $11.6\pm0.8$, $8.7\pm0.81$, and $4.7\pm0.6$ Jy,
  respectively. The median intensities in an annulus surrounding the
  aperture up to the radius $50\arcsec$ was used in the background
  subtraction. The \textit{Spitzer}/MIPS 24- and 70-$\mu$m flux
densities, $3.0\pm0.4$ and $349\pm 44$ mJy, were adopted from
  \citet{2013A&A...557A..98T}. The \textit{Herschel} flux densities
are smaller than those listed for Cha-MMS1 in
\cite{2012A&A...545A.145W}, probably because we have used a smaller
aperture (the apertures are not given in the paper of Winston et al.).
 
\begin{figure}[!ht]
  \centering
  \resizebox{0.95\hsize}{!}{\includegraphics{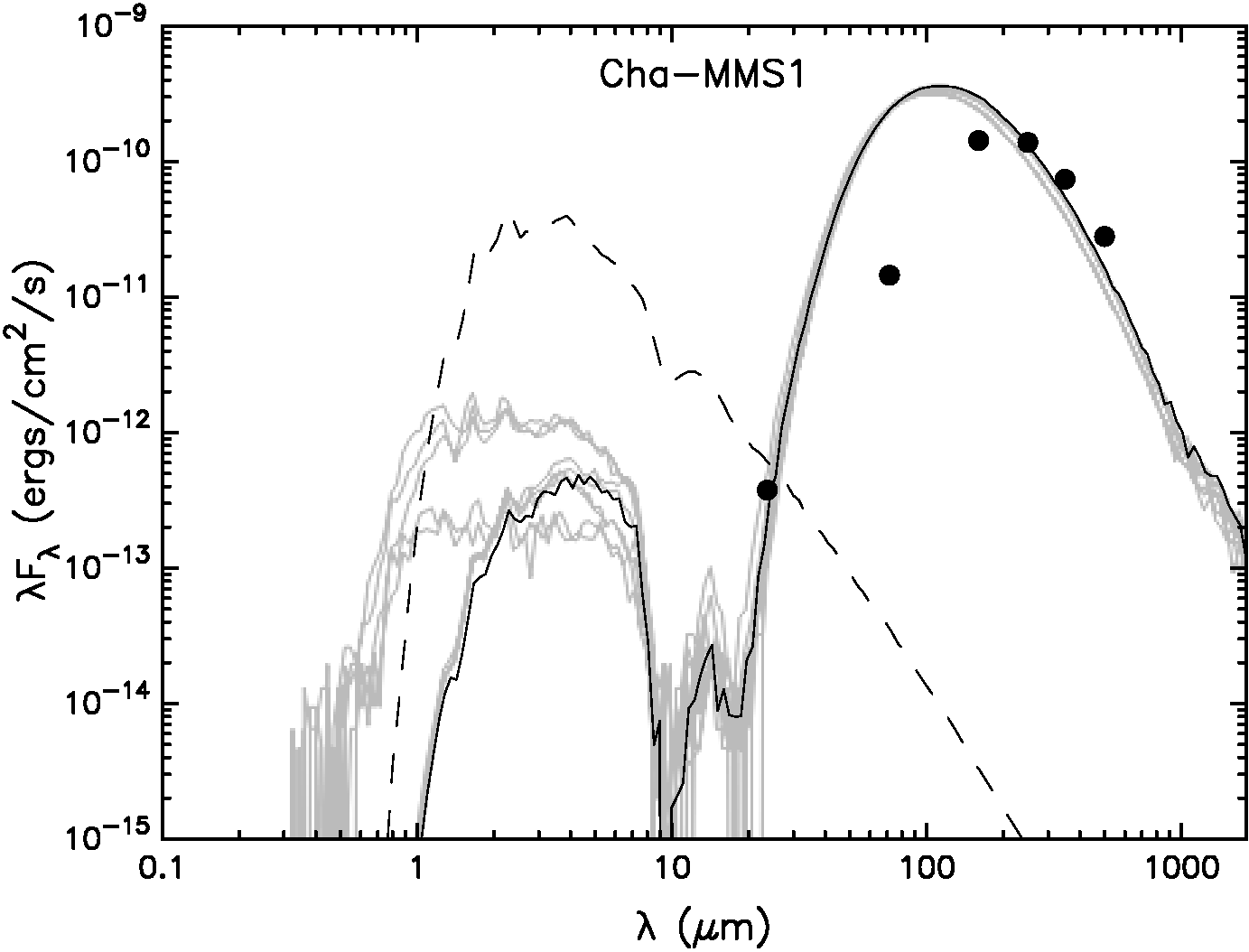}}
  \caption{Spectral energy distribution for Cha-MMS1. The observed
    flux density values are shown as filled circles. The
      error bars are smaller than the data points in the
      figure. Ten best model fits are shown: the solid black line
    shows the best-fitting model, and the grey lines illustrate the
    subsequent nine best-fit models. The dashed line shows the SED of
    the stellar photosphere corresponding to the central source in the
    best-fitting model (in the absence of circumstellar material).}
\label{figure:sed}
\end{figure}

The SED of Cha-MMS1 was analysed by fitting model SEDs to the observed
flux densities from a large pre-computed grid of YSO models
\citep{2006ApJS..167..256R, 2007ApJS..169..328R}\footnote{The online
  SED-fitting tool is available at {\tt
    http://caravan.astro.wisc.edu/protostars/}}.  The grid consists of
20\,000 two-dimensional YSO radiation transfer models covering a wide
range of stellar masses (0.1--50 M$_{\sun}$) and evolutionary stages.
Each model outputs SEDs at 10 different viewing angles (inclinations)
ranging from edge-on ($90\degr$) to pole-on ($0\degr$, resulting in a
total of 200\,000 SEDs. The model fitter requires a distance range
within which to fit the observed data.  We assumed that the source
distance is in the range 135--165 pc \citep{1997A&A...327.1194W}. The
other assumption we made was that the foreground interstellar
extinction is no more than $A_{\rm V}=20$.  In Fig.~\ref{figure:sed}
we show the SED for Cha-MMS1 along with the best-fitting model and the
subsequent nine best-fit models (shown as grey lines).

The models could not fit the 70 $\mu$m data point well.
We note that inclusion of the 70 $\mu$m data point icreases
significantly (by a factor of $\sim9$) the $\chi_{\rm best}^2/N_{\rm
  data}$ value ($N_{\rm data}$ is the number of data points) with
respect to the calculation where the 70 $\mu$m flux is not used in the fit. 
Moreover, the models predict slightly higher 160 $\mu$m
flux densities than the observed value.
The resulting model parameters are given in Table~\ref{table:SED}.  In
addition to the best-fit model, we show the range of possible
parameter values that can be derived from the subsequent nine best
models.

The best-fit model stellar mass and total luminosity are found to be
$M_{\star}\sim0.12$ M$_{\sun}$ and $L_{\rm tot}\sim0.6$ L$_{\sun}$.
We note that the latter includes contributions from the central star
and accretion, assuming that all the accretion energy is radiated
away. Moreover, it is corrected for foreground extinction and does not
depend on the viewing angle. The envelope mass, $M_{\rm env}\sim0.9$
M$_{\sun}$, determined from the long wavelength emission, is
about half of the value 2.2 M$_{\sun}$ derived from NH$_3$, which is a
reasonable agreement given the assumptions used in the analysis
[e.g., $X({\rm NH_3})$].

\citet{2006ApJS..167..256R} defined different stages of evolution
based on the value of the stellar mass, $M_{\star}$, and the envelope
accretion rate, $\dot{M}_{\rm env}$. For Cha-MMS1, we derive the
$\dot{M}_{\rm env}/M_{\star}$ ratio $\sim 4\times10^{-4}$ yr$^{-1}$ which
in this classification scheme suggests that it represents a deeply
embedded Stage 0 source ($\dot{M}_{\rm env}/M_{\star}\gg10^{-6}$
yr$^{-1}$) seen nearly edge-on (the inclination angle is
$\sim 70\degr$).  This early evolutionary stage is also cha\-racterised by
an envelope whose mass is $M_{\rm env}\gg M_{\star}$ and $M_{\rm
  disk}$.  The derived luminosity, 0.6 L$_{\sun}$, is higher than the
value expected for a candidate FHSC ($<0.1$
L$_{\sun}$; e.g., \citealt{2010ApJ...722L..33E};
\citealt{2011ApJ...743..201P}; \citealt{2012ApJ...751...89C}).  

Our results suggest that if the central source is a true protostar, it
represents a Class 0 object, an accreting second hydrostatic core at a
very early stage of evolution. It should be noted, that the models of
\citet{2006ApJS..167..256R} only cover YSOs with central temperatures above 
2\,000 K. The model fails to fit the 70 $\mu$ flux density which has been found to correlate closely with the internal luminosity of the protostar 
\citep{2008ApJS..179..249D}. Based on the correlation found by \cite{2008ApJS..179..249D} and on the predictions of \citet{2012A&A...545A..98C} for the FHSC flux densities at 24 and 70 $\mu$m, \citet{2013A&A...557A..98T} concluded that 
the central source of Cha-MMS1 has a very low luminosity, $\sim 0.1 L_\odot$, 
and it is likely to be a FHSC. On the other hand, the models of 
\citet{2012A&A...545A..98C} suggest that it is very
difficult to distinguish between embedded Class 0 protostars and FHSCs
based solely on SED information.

\begin{table}
\renewcommand{\footnoterule}{}
\caption{Results of SED modelling.}
{\tiny
\begin{minipage}{1\columnwidth}
\centering
\label{table:SED}
\begin{tabular}{c c}
\hline\hline
Parameter\tablefootmark{a} & Value\tablefootmark{b}\\
\hline
$\chi^2/N_{\rm data}$ & 85.34 [86.29--95.98] \\
Inclination, $i$ [$\degr$] & 69.51 [63.26--87.13]\\
Stellar age, $\tau_{\star}$ [$10^4$ yr] & 2.65 [2.65--6.95] \\
Stellar mass, $M_{\star}$ [M$_{\sun}$] & 0.12 [0.12--0.18] \\
Stellar temperature, $T_{\star}$ [$10^3$ K] & 2.78 [2.78--3.02]\\
Envelope accretion rate, $\dot{M}_{\rm env}$ [$10^{-5}$ M$_{\sun}$~yr$^{-1}$] &
 4.26 [2.84--4.91]\\
Disk mass, $M_{\rm disk}$ [M$_{\sun}$] & 0.0013 [$1.7\times10^{-4}$--0.0095]\\
Total luminosity, $L_{\rm tot}$ [L$_{\sun}$] & 0.59 [0.43--0.68]\\
Envelope mass, $M_{\rm env}$ [M$_{\sun}$] & 0.88 [0.46--0.88] \\
Stage & 0 \\
\hline
\end{tabular}
\tablefoot{\tablefoottext{a}{The parameters are described in detail in
Robitaille et al. (2006).}\tablefoottext{b}{For each model parameter, the
best-fit value is given, followed by a range defined by the minimum and
maximum values obtained from the subsequent nine best-fit models.} }
\end{minipage} }
\end{table}

\section{Discussion}

\label{discussion}

\subsection{The high column density ammonia core}

One important conclusion from this work is that with ATCA, we
  detect a compact high column density structure in the NH$_3(1,1)$
  line which is ``invisible'' at single dish (Parkes) resolution.  The
  average {\sl para}-ammonia column density which we derive ($\sim
  1\times10^{15}$ cm$^{-2}$) is likely an underestimate due to high
  optical depth in the satellite transitions.  Such high column
  densities of ammonia in cold cores are very rare and, actually, we
  are not aware of any comparable examples.  For example, towards L1544,
  \cite{2007A&A...470..221C} found a total ({\sl ortho} + {\sl para})
  NH$_{3}$ column density of $7\times10^{14}$ cm$^{-2}$ using the VLA.
  This value is derived assuming thermal relative populations of the
  {\sl ortho} and {\sl para} states, impying that the {\sl
    para}-ammonia column density in L1544 is $\sim1\times10^{14}$
  cm$^{-2}$. This is an order of magnitude less than our value in
  Cha-MMS1.

One can check this by comparing the 870 $\mu$m map of Cha-MMS1 from
\cite{2011A&A...527A.145B} (see also \citealt{2013A&A...557A..98T})
made with the LABOCA receiver on the APEX telescope (HPBW $21
\arcsec$) with our NH$_{3}$ column density map. This is shown in
Fig.~\ref{figure:laboca} where we have smoothed the NH$_{3}$ to the
dust continuum resolution. One sees that the dust continuum and the
ammonia column trace somewhat different regions and in fact the
ammonia peak is $25 \arcsec$ to the south of the 870 $\mu$m peak.
This may partly reflect the ATCA missing flux problem
discussed in Sect.~2.2.  However, it might also indicate a gradient
in the ammonia abundance.  We have attempted to estimate the ammonia
abundance in this area assuming a dust temperature of 12.5 K and a
a dust opacity of 0.0083 cm$^{2}$g$^{-1}$  
at $\lambda=870\,\mu$m (which is consistent with the cross-section at 
 $\lambda=250\,\mu$m assumed in Sect.~2.3).  We
derive a mean H$_{2}$ column density over the ammonia source of
$7.6\times10^{22}$ cm$^{-2}$ and a mean total ammonia abundance
[NH$_{3}$]/[H$_{2}$] = $3 \times 10^{-8}$ (assuming {\sl ortho}/{\sl
  para} $\sim 1$).  The dust continuum angular resolution does not
suffice to make a reliable estimate of the abundance gradient but a
difference of a factor 2 from north to south seems plausible.
Clearly, higher angular resolution observations of this region in a
variety of molecular species and in particular $^{15}$NH$_{3}$ are
needed to clarify this situation.

The mean NH$_{3}$ abundance is quite compatible with past ammonia
abundance determinations in cores (e.g., \citealt{1993A&AS...98...51H}; 
\citealt{2011ApJ...741..110D}) but applies to a region considerably
denser than in most previous work ($10^6$ cm$^{-3}$, see
Table~\ref{table:parameters}).  The timescale for freeze-out of gas
phase ammonia on dust grain surfaces at this density is of order
$10^4$ years suggesting that the age of the high-density ammonia core
cannot be much greater than this.  In general, the ammonia abundance
reflects the molecular nitrogen abundance and ammonia correlates well
with N$_{2}$H$^{+}$.  This suggests that ALMA observations of
N$_{2}$H$^{+}$ would be a fruitful way of studying the kinematics of
the core surrounding Cha-MMS1.

\subsection {Core kinematics}

The kinematics of the molecular cloud in the neighbourhood of Cha-MMS1
are complex as one realises studying the ${\rm N_2H^+}(1-0)$ map of
\cite{2011ApJ...743..108L} which covers an area of roughly $200
\arcsec$ in diameter with an angular resolution of $36 \arcsec$.  Our
results can also be compared with the data of \cite{2013A&A...557A..98T}
who present in several tracers the results of cuts parallel and
perpendicular to the filamentary structure seen in the APEX LABOCA map
(see \citealt{2013A&A...557A..98T}, Fig.1a).  Our ATCA ammonia map
covers a  more limited region with a higher angular resolution
than in the above mentioned studies (see Table~1) and traces the kinematics 
of high density [$n({\rm H_2})\, = \, 10^6$ cm$^{-3}$] material. A general aim of
such observations is to compare the kinematics of the high density
core nucleus with that of the surrounding lower density material.  One
can expect to find evidence for outflows or accretion associated with
embedded protostars as well as for rotation.  In the latter case, a
naive expectation is that due to angular momentum conservation, the
rotation axis is the same for high and low density gas but that
angular velocity increases rapidly on small size scales.

We first note that the direction of the velocity gradient found by us
is quite different from that found by Tsitali et al. perpendicular to
the "filamentary structure" seen in the LABOCA map.  The reason for
that may lie in the positioning of the cuts, angular resolution, and
the molecular probes used by Tsitali et al. Our radial velocities
increase from the northeast to the south in the same way as in the
single-dish ${\rm N_2H^+}$ map of Ladd et al., but the gradient is
steeper in the present NH$_3$ map.  This suggests differential
rotation with the angular ve\-lo\-ci\-ty increasing towards the
rotation axis.  The axis of rotation has a position angle of
$130\degr$, and is roughly delineated in the velocity channel centred
at 4.5 km~s$^{-1}$ in Fig. \ref{figure:channelmap}.  The steepness of
the velocity gradient corroborates the notion that the ammonia
emission traces the central parts of the core.

The bright N-S oriented ridge seen in the integrated intensity map
shares the overall velocity gradient of the central region.  The
largest diameter of the ridge is about 5\,000 AU, and its symmetry
axis is tilted with respect to the overall rotation axis by
$45\degr$. This configuration resembles very much the structure
observed around the YSO Barnard 1c (\citealt{2006ApJ...652.1374M}, see
their Fig.8). Unlike Cha-MMS1, Barnard 1c reached an advanced stage with 
well-developed molecular outflow.
The column density map (Fig.~\ref{figure:N11map}) shows that the 
largest ammonia column densities are found in two separate regions, 
near the northern and southern ends of the ridge.  
The two most prominent column density maxima lie in the vicinity of 
the rotation axis, some $30\arcsec$ ($\sim 5000$ AU) apart.  
The northern ammonia maximum lies close to the far-infrared peak 
determining the location of the H$_2$ column density maximum.
As discussed above, the southern ammonia maximum is likely to trace a 
local enhancement in the fractional ammonia abundance.

Both \citet{2007ApJ...664..964H} and \citet{2011ApJ...743..108L}
suggest that outflow from the neighbouring Class I protostar IRS4,
lying about $80\arcsec$ ($\sim$13\,000 AU, 0.06 pc) NE from our map 
centre, is interacting with the Cha-MMS1 core. 
According to both studies the momentum of the IRS4 outflow is large 
enough to affect the core dynamics. 
On the other hand, \citet{2011ApJ...743..108L} find
that the outflow might be deflected from the Cha-MMS1 core, and they 
suggest that the outflow is not colliding with the core directly but 
delivers ``a glancing blow''. 

While it is possible that the outflow from IRS4 has influenced the
velocity structure in the northeastern side of the core, and may have
contributed to the core collapse through compression it is less likely
that the overall velocity gradient could have been directly caused by
external momentum input.  An outflow cannot easily exert torque on a
gaseous body; more likely it will plough its way through the cloud
\citep[see][Sect. 4.2.1]{2011ApJ...743..108L}. It is therefore
difficult to explain the redshifted gas in the southern part by the
outflow from IRS4.  The most believable explanation for the large
velocity gradient is therefore spin-up resulting from (partial)
conservation of the angular momentum during the core collapse. 
  Another possibility however is that infall onto the Cha-MMS1
  \textit{Spitzer} source is responsible for the high velocity
  gradient close to the protostar.  \cite{2012ApJ...748...16T}, in
  their study of Class 0 protostellar envelopes, point out that the
  kinematic signatures of infall and rotation are similar and hence
  difficult to distinguish.  Therefore, it is certainly possible that
  the real situation is a combination of the two.

Are there any signs of a molecular outflow along the rotational axis
of the core  as in the case of Barnard 1c?  
An elongated, hourglass-shaped structure in the likeness of some 
previously observed outflow lobes is seen in the channel maps, albeit 
in a very narrow LSR velocity range of 4.4-4.5 \kmps 
(see Fig.~\ref{figure:channelmap}).
The structure is aligned with the approximately northwest-southeast 
oriented rotation axis (P.A. $130\degr$) and seems to protrude from 
the densest part of the core on its southeastern side.  
A symmetric, elongated ammonia structure and the absence of 
high-velocity wings could possibly be explained by a dense outflow or 
by compressed ouflow cavity walls seen nearly side on.  

Ammonia has been previously observed to probe cavity walls around
evolved molecular outflows (\citealt{1995ApJ...443..682A};
\citealt{2008A&A...485..517B}), and to take part in collimated
high-velocity outflows (\citealt{1993ApJ...417L..45B};
\citealt{1995ApJ...443L..37T}; \citealt{1996ApJ...456..677C}). 
In the latter case, the gas properties (line shapes, kinetic
temperatures, enhanced ammonia abundances) show clear evidence for
shocks.  In the present ammonia data no sign of shocked gas is,
however, seen. The ammonia linewidths (see
Fig.~\ref{figure:widthmap}) are below 0.3 \kmps in the southeastern part of the
putative NW-SE ``outflow'', but increase to slightly above 0.3 \kmps
in a conical region in the northwest. The analysis of the
NH$_3$ profiles suggests the presence of a warm gas component in the
neighbourhood of the \textit{Spitzer} 24-$\mu$m source and the rotation axis: A
two-layer model with a cool foreground component, and an underlying
dense or/and warm component is applicable to this elongated region.
The distribution of this component overlaps largely
with the region showing 1.2 cm continuum emission, and wide-band
  ${\rm NH_3}(2,2)$ emission. This coincidence suggests that the
present ammonia map partly probes gas heated by an outflow embedded in
Cha-MMS1.

Another intriguing stucture reminiscent of an outflow is seen in the 
wide-band $(1,1)$ map (Fig.~\ref{figure:widebandimages}): three
maxima, the brightest one associated with the NH$_3$ column density 
maximum near the far-IR peak, and two others on its southernside. 
These features lie along the bright ridge seen in the integrated line 
emission map (P.A. $175\degr$). 
The two subsidiary peaks have slightly arc-like shapes suggesting 
shock compression. 
Also in these positions ammonia lines are narrow ($0.2-0.3$ \kmps), 
and no trace of pedestals or wings can be seen in the $(1,1)$ 
high-resolution spectra. 
The wiggling secondary maximum in the (1,1) column density map 
mentioned above is associated with the southernmost wide-band $(1,1)$ 
peak.

The spectra shown in Fig.~\ref{figure:ATCA_spectra} as well as 
other spectra in the vicinity of the \textit{Spitzer} 24-$\mu$m source show no
clear evidence for accretion.  In fact absorption in this region is
{\sl blue-shifted} with respect to the embedded warmer gas. We note
here however that it is also possible that this "self absorption"
occurs in the foreground more extended component which is mainly
responsible for the lower optical depth emission seen with single dish
telescopes.

\subsection{On the nature of Cha-MMS1}

The SED described in Sect.~\ref{Results:Sed} suggests that, if
Cha-MMS1 has entered into a protostellar stage, it is still a Class 0
-type object.  Class 0 protostars are known to drive outflows
\citep[e.g.][]{2006ApJ...646.1070A}, and in such of an object a
strongly collimated high-velocity jet can act as the driving force
behind the slower velocity molecular outflow.  (e.g.,
\citealt{2007prpl.conf..245A}). In Cha-MMS1, emission at visual and IR
wavelengths from a possible high-velocity jet could be totally
obscured by the dense core.  The presence of a jet is suggested by the
tentative detection of radio continuum emission
(Sect. \ref{widebandimages}), which traces free-free emission from
shock-ionised gas.  The bending of this feature towards north does not
agree with the linearity of the NW-SE oriented ammonia structure.  On
the other hand, three of the continuum peaks lie close to the
axis defined by the wide band $(1,1)$ maxima
(Fig.~\ref{figure:widebandimages}).  The total 1.3 cm flux density of
the continuum feature seen in Fig.~\ref{figure:widebandimages} is
$0.8\pm0.4$ mJy.  Assuming a spectral index of $\alpha=1.2$ (see
Sect.~\ref{widebandimages}), and using the relationship between the
outflow momentum rate and the 6\,cm radio continuum luminosity derived
by \cite{1998AJ....116.2953A} we obtain an upper limit for the
momentum rate of $\dot{P}=5\times10^{-6}\,M_{\odot} {\rm
  yr^{-1}\,km\,s^{-1}}$.  Using the stellar age $\sim 3\times10^4$
  yr derived from the SED modelling (Sect.~\ref{Results:Sed}), and
  adopting a velocity of $30\,{\rm km\,s^{-1}}$ for the putative
  molecular outflow, the mass of the outflow becomes
  $0.005\,M_{\odot}$.  In the situation where the described outflow
is seen side-on, only a fraction of it is seen within a telescope
beam, and we estimate that the integrated intensity of the CO$(J=3-2)$
line as observed with a 10-m telescope (like ASTE) would be $\int\,
T_{\rm MB} {\rm d}\upsilon \sim 0.5 \, {\rm K\,km\,s^{-1}}$, and might
remain undetected taking into account blending with the quiescent gas
and the outflow from IRS4 (see Fig.~7 of
\citealt{2007ApJ...664..964H}). On the other hand, in the case of a
small inclination angle the same outflow should produce a signal
detectable with single-dish telescopes.

Two principal mechanisms have been proposed to drive protostellar
jets. Both are based on the magnetocentrifugal mechanism
(\citealt{1982MNRAS.199..883B}; \citealt{1994ApJ...429..781S}) but
related to different driving sources of the outflow.  Firstly, the
magnetocentrifugal mechanism can be operational through an accretion
disk alone (e.g., \citealt{1982MNRAS.199..883B},
\citealt{2007prpl.conf..277P}), assuming that the magnetic field of
the disk is twisted enough to form a sufficiently open poloidal field
geometry, through which excess matter and angular momentum can escape
the system (the disk-wind model). Secondly, the magnetocentrifugal
mechanism can also be active in a system consisting of an accretion
disk and a protostar (\citealt{1994ApJ...429..781S}). The rotational
interaction between the magnetic fields of a disk and a young star
creates open field lines near the equator of the system, where matter
is blown out as bipolar jets (the X-wind model).  These mechanisms
might be relevant in an early stage already. Several numerical MHD
models \citep{1998ApJ...502L.163T, 2000ApJ...528L..41T,
  2002ApJ...575..306T,2006ApJ...641..949B, 2008ApJ...676.1088M} have
also shown that a FHSC can produce low-velocity outflows.  In these
simulations, FHSC outflows have wide opening angles and extend to a
distance of a few times the width of the outflow. However, according
to the recent model calculations of \cite{2012MNRAS.423L..45P},
low-mass FHSCs can generate well-collimated, slow ($\sim 2-7$ \kmps)
outflows, expanding to several thousand AU.

The projected total length of the hypothetical NW-SE outflow is (based 
on the structure in Fig.8) approximately $9\,000$ AU. 
Adopting the range of the characteristic outflow speeds from 
\cite{2012MNRAS.423L..45P},
this length corresponds to ages between $3\,000$ and $11\,000$ yr. 
These times are slightly longer than the expected lifetimes of FHSCs
\citep{2012A&A...545A..98C},
but comparable to the dynamical time-scale of the outflow associated 
with the candidate FHSC Per-Bolo 58 ($\upsilon = 2.4$ \kmps, 
$\tau_{\rm d} = 16\,000$ yr; \citealt{2011ApJ...742....1D}).

There are a number of molecular outflows observed from very young
objects.  Some of them are suspected to be still at the first core
stage and some are likely to be more advanced very low luminosity
objects \citep[VeLLOs; see][]{2008ApJS..179..249D}.  Examples of this
are L673-7 \citep{2010ApJ...721..995D}, Per-Bolo 58
(\citealt{2010ApJ...722L..33E}; \citealt{2011ApJ...742....1D}), L1448 IRS 2E
\citep{2010ApJ...715.1344C}; L1451 \citep{2011ApJ...743..201P}, CB 17
MMS \citep{2012ApJ...751...89C}, and L1521F-IRS \citep{2013ApJ...774...20T}.
In general, the resolution and the noise level in the observations
towards FHSC candidates are not good enough to exclude a
low-luminosity protostar.  L1521F-IRS deserves a particular mention as
Takahashi et al. (2013) state that they can see two distinct velocity
components in their observations: one corresponding to wide slow
outflow and more highly collimated fast outflow. They suspect that the
slow component originates from the FHSC stage.

The theoretical and observational studies described above show that
outflows are characteristic of very young protostars.  Several
features of Cha-MMS1: a rotating core with an embedded 24-$\mu$m
continuum source, the tentative detection of radio continuum, the
hourglass-shaped ammonia distribution at the nominal velocity of the
cloud, suggest the presence of a young molecular outflow.  However, in
contrast to the VeLLO and candidate FHSC outflows mentioned above, no
CO outflow originating in Cha-MMS1 has been detected. If such an
outflow exists, the high-velocity wings may still be confined in the
very inner regions of the core, and can have escaped the previous
single-dish surveys. To confirm or disprove this possibility more
detailed and sensitive observations than available so far are needed.
High-resolution submillimetre/far-IR continuum observations could be
used to study the shape of the compact far-IR source, which may be
associated with the FHSC or a protostellar disk.  Also, line
observations in high-density tracer molecules could help us to
understand the velocity structure of the core in more detail.
Finally, deep radio con\-ti\-nuum observations can confirm the
presence of free-free emission from ionised jets.

The rotational speed ($\Omega\sim6.5\times10^{-13}$ s$^{-1}$) and the
ratio of rotational and gravitational energies ($\sim 7 \%$) derived
for the core are large, and rotation is probably influencing strongly
the star formation process in Cha-MMS1.  Furthermore, the region has
been reported to be significantly magnetised, with the large-scale
magnetic field oriented roughly in the same way as the rotation axis
inferred from the present ammonia data (\citealt{1994MNRAS.268....1W};
\citealt{2005A&A...431..149H}, their Fig.8).  The combination of fast
rotation and magnetic fields should have resulted in the formation of
a disk-like first core, probably generating outflows through toroidal
magnetic pressure (\citealt{1998ApJ...502L.163T};
\citealt{2004MNRAS.348L...1M}; \citealt{2006ApJ...641..949B}).

According to MHD models, the circumstellar disk formed from the first
core should be small compared with that in an unmagnetised core
because of the removal of the angular momentum by magnetic braking.
Nevertheless, a high rotation speed can have led to the fragmentation
of the disk and the formation of a protobinary or a multiple object,
at a phase (density) determined by the ratio of the rotational and
magnetic energies (\citealt{2008ApJ...677..327M};
\citealt{2011MNRAS.413.2767M}; \citealt{2013A&A...554A..17J}).  
The ammonia data show indications of outflows along two axes with
  the position angles $130\degr$ and $175\degr$ (see Sect.~4.2). Jet
  precession or a double jet associated with a binary would possibly
  explain these features.

\section{Conclusions}

\label{conclusions}

We have mapped the Cha-MMS1 core in the NH$_3(1,1)$ line and in the
1.2 cm continuum using the Australia Telescope Compact Array at
spatial resolution of about 1\,000 AU. The observations reveal a
high-column density ammonia core with a steep (20 \kmps pc$^{-1}$)
velocity gradient. The gradient can be interpreted as rotation around
a NW-SE oriented axis which passes through the position of a
\textit{Spitzer} 24-$\mu$m point source. The ammonia spectra are
self-absorbed near the rotation axis. A two-layer model used to fit
the absorption suggests that the underlying component has an elevated
temperature near the \textit{Spitzer} source.  A string of weak 1.2 cm
continuum sources are tentatively detected in this region.

It seems natural to explain the presence of warm gas close to the
rotation axis in terms heating by an embedded outflow. However, no evidence for
high-velocity gas or shocks are found in the spectral lines
observed towards this core, so the question about
the evolutionary stage of the central source remains open. An analysis
of the SED using the models of \cite{2006ApJS..167..256R} suggests
that if the central source is a YSO it belongs to the Class 0. On the
other hand, as pointed out by \cite{2013A&A...557A..98T}, the
$70\,\mu$m flux density of the central source implies a very low
luminosity which would be consistent with an FHSC, but clearly too low
for a true protostar. 

Observations at a high spatial resolution of $\sim 100$ AU ($\leq 1\arcsec$)
will probably clarify this issue. Besides a search for a compact
outflow in CO, probing the distribution and kinematics of dense gas in the
neighourhood of the \textit{Spitzer} source would be useful. The
observed ammonia lines have very large optical thicknesses, and it
is likely that the fractional abundances of both ${\rm NH_3}$ and
${\rm N_2H^+}$ are high in the centre of the core. These molecules, and 
their $^{15}$N substituted variants look therefore suitable for probing the 
the interior parts of the core.

\begin{acknowledgements}
  We wish to thank the ATCA staff for their help during our
  observations. Financial support from the Academy of Finland projects
  132291, 141017, and 218159, from the research programme ``Active
  Suns'' at the University of Helsinki, and from a Jenny and Antti
  Wihuri Foundation grant are acknowledged.\\

\indent This work is based in part on observations made with the
\textit{Spitzer Space Telescope}, which is operated by the Jet Propulsion
Laboratory, California Institute of Technology under a contract with NASA.\\

\indent SPIRE has been developed by a consortium of institutes led by Cardiff
Univ.  (UK) with Univ. Lethbridge (Canada); NAOC (China); CEA, LAM
(France); IFSI, Univ. Padua (Italy); IAC (Spain); Stockholm
Observatory (Sweden); Imperial College London, RAL, UCL-MSSL, UKATC,
Univ. Sussex (UK); Caltech, JPL, NHSC, Univ. Colorado (USA). This
development has been supported by national funding agencies: CSA
(Canada);NAOC (China); CEA, CNES, CNRS (France); ASI (Italy); MCINN
(Spain); SNSB (Sweden); STFC (UK); and NASA (USA). PACS has been
developed by a consortium of institutes led by MPE (Germany) with UVIE
(Austria); KU Leuven, CSL, IMEC (Belgium); CEA, LAM(France); MPIA
(Germany); INAFIFSI/ OAA/OAP/OAT, LENS, SISSA (Italy); IAC
(Spain). This development has been supported by the funding agencies
BMVIT (Austria), ESA-PRODEX (Belgium), CEA/CNES (France), DLR
(Germany), ASI/INAF (Italy), and CICYT/MCYT (Spain).\\

\indent This research has made use of NASA's Astrophysics Data System.

\end{acknowledgements}

\bibliographystyle{aa}

\bibliography{Cha_refs}

\end{document}